\documentclass[aps,prd,reprint,floatfix,notitlepage,a4paper]{revtex4-2}
\usepackage[T1]{fontenc}
\usepackage{lmodern}
\usepackage{amsmath,amssymb,mathtools,bm,mathrsfs}
\usepackage{slashed}
\usepackage{microtype,booktabs,array,graphicx}
\usepackage{capt-of}
\usepackage{placeins}
\usepackage{xcolor}
\usepackage{hyperref}
\hypersetup{
 colorlinks=true,
 linkcolor=blue,
 citecolor=blue,
 urlcolor=blue,
 pdfborder={0 0 0}
}
\usepackage{orcidlink}
\usepackage{tikz}
\usetikzlibrary{decorations.pathmorphing,decorations.markings,arrows.meta}
\tikzset{
 gluon/.style={draw,decorate,decoration={coil,aspect=.75,segment length=4.pt,amplitude=2.5pt,pre length=2.5pt,post length=0pt},line width=.72pt},
 gluonloop/.style={draw,decorate,decoration={coil,aspect=.5,segment length=4pt,amplitude=2.5pt,pre length=0pt,post length=0pt},line width=.92pt},
 gluonloopinner/.style={draw,decorate,decoration={coil,aspect=.75,segment length=2.4pt,amplitude=-1.55pt,pre length=0pt,post length=0pt},line width=.72pt},
 gline/.style={gluon},
 qline/.style={draw,line width=.82pt},
 ghostline/.style={dashed,line width=.65pt},
 source/.style={dashed,line width=.65pt}
}
\allowdisplaybreaks[2]
\newcommand{\tr}{\operatorname{tr}}
\newcommand{\dd}{\mathrm{d}}
\newcommand{\CA}{C_A}
\newcommand{\CF}{C_F}
\newcommand{\TR}{T_R}
\newcommand{\TF}{T_F}
\newcommand{\sumint}{\mathop{\sum\mkern-24mu\int}}
\newcommand{\Db}{\Delta_B}
\begin{document}

\title{Magnetic-field dependence of the QCD electric screening mass parameter through $g^4$ in hot and dense matter}
\author{Najmul Haque\orcidlink{0000-0001-6448-089X}}
\email{nhaque@niser.ac.in}
\affiliation{School of Physical Sciences, National Institute of Science Education and Research, An OCC of Homi Bhabha National Institute,\\  Jatni, Khurda 752050, India}

\begin{abstract}
A constant magnetic field modifies the hard electric screening sector of hot
and dense QCD and generates a directional splitting at $\mathcal{O}(g^4)$. We determine the strict-hard screening pole through $g^4$ for three massless flavors with physical $u,d,s$ charges, retaining the full $eB/T^2$ dependence without a weak- or strong-field expansion. The shifted-frequency construction used for the two-loop exchange contribution is a background-field Feynman-gauge curvature; we derive and apply its local conversion to the ordinary Feynman-gauge self-energy before assembling the screening pole. The leading magnetic shift is isotropic, while the $g^4$ directional splitting arises from the one-loop anisotropic kinetic normalization; the symmetry-allowed finite-density $A_0F_{12}$ mixing is numerically small over the studied range for physical $u,d,s$ charges. For $\mu/T=0$ and $1$, the magnetic $g^4$ coefficient is positive over $0\le eB/T^2\le4$ and is strongly reduced by density. Extending the calculation to $0\le\mu/T\le2$ at $eB/T^2=1$ and $4$ reveals sign reversals of both the leading magnetic coefficient and the directional splitting at sufficiently large density. The zero-field limit reproduces the Braaten-Nieto result and its finite-density extension by Vuorinen. The running coupling used only for the phenomenological magnetic mass shift is chosen independently of $B$, so that all magnetic-field dependence through $g^4$ resides in the calculated coefficients.
\end{abstract}
\maketitle

\section{Introduction}
The static electric sector of a quark-gluon plasma is screened on a
length scale of order $(gT)^{-1}$. In non-Abelian thermal field theory
it is important to distinguish the hard electric mass parameter $m_E$,
obtained by matching modes of order $T$ onto the electrostatic effective
theory, from the physical long-distance Debye screening mass. Beyond
leading order these are not the same quantity. At
zero magnetic field, the Braaten-Nieto strict matching calculation determines the hard electrostatic mass parameter through $\mathcal{O}(g^4)$ at zero density~\cite{Braaten:1995jr}, while Vuorinen extended the same coefficient to finite quark chemical potentials~\cite{Vuorinen:2003fs}. More generally, dimensional reduction organizes the hard scale by matching static Green functions onto a three-dimensional bosonic effective theory~\cite{Kajantie:1995dw}; higher-order matching of other EQCD parameters, such as the effective gauge coupling, has been carried out in Ref.~\cite{Laine:2005ai}. The resummed screening pole additionally contains the well-known soft contribution that starts at $\mathcal{O}(g^3T^2)$ in the
squared mass~\cite{Rebhan:1993az}. Perturbative studies of static electromagnetic screening in relativistic plasmas provide useful historical context~\cite{Kapusta:1992fm,Blaizot:1995kg}. In the non-Abelian case, the long-distance Debye screening length beyond leading order is sensitive to the
soft sector and admits a gauge-invariant nonperturbative definition~\cite{Arnold:1995bh}. The hard Debye parameter itself has subsequently been pushed to three-loop order in hot Yang-Mills theory~\cite{Ghisoiu:2015uza}.

At leading order the Debye mass sets the inverse color-electric screening
length, so electric screening also controls the long-distance part of the
in-medium interaction between a heavy quark and antiquark. This connection
underlies the classic Matsui-Satz proposal that sufficiently strong color
screening in a quark-gluon plasma can suppress quarkonium binding
~\cite{Matsui:1986dk}. Screened-potential studies make the relation to the
screening length explicit by determining when individual heavy-quark bound
states cease to exist~\cite{Karsch:1987pv}. Strong transient magnetic fields
are generated in noncentral relativistic heavy-ion collisions by the rapidly
moving charges of the colliding ions~\cite{Skokov:2009qp,Deng:2012pc,Tuchin:2013ie}.
These fields motivate studying how a magnetic background modifies the hard
QCD electric screening parameter $m_E$ in hot and dense matter.
An external magnetic field introduces a second hard scale and
breaks spatial rotational symmetry. The resulting longitudinal-transverse
pressure anisotropy of strongly magnetized deconfined QCD matter has been
examined at one loop~\cite{Karmakar:2019tdp}. Quarks are Landau quantized,
whereas the gluons remain electromagnetically neutral. Consequently the
zero-momentum one-loop temporal self-energy coefficient remains a scalar,
but the $\mathcal{O}(k^2)$ part of the static temporal polarization separates into longitudinal and transverse coefficients. The general tensor structure of
gauge-boson propagators and their collective modes in hot magnetized matter
has been studied within HTL theory~\cite{Karmakar:2018aig}. Direction-dependent screening is familiar more generally in anisotropic QED
and QCD plasmas~\cite{Birse:2003qp}. Related magnetic-field-induced anisotropy of the static $\overline{Q}Q$ potential has been observed directly in lattice QCD~\cite{Bonati:2014ksa}, while electric and magnetic screening masses in strong external magnetic fields have also been studied on the lattice~\cite{Bonati:2017uvz}. The anisotropic complex heavy-quarkonium potential at arbitrary magnetic-field strength has also been studied using the one-loop gluon polarization tensor and the corresponding dielectric response
~\cite{Sebastian:2023tlw}. At two loops, the field dependence resides in quark-bearing subgraphs, and a consistent calculation must keep the full Landau-level or proper-time structure rather than insert a weak-field propagator into a zero-field final formula. Related exact Landau-level representations of the one-loop polarization tensor can be found in Ref.~\cite{Ayala:2019akk}. In thermal QED, Alexandre evaluated the one-loop photon polarization tensor in a homogeneous magnetic field using Schwinger's proper-time formalism and discussed its implications for Debye screening~\cite{Alexandre:2000jc}. One-loop HTL analyses of magnetized hot and dense QCD also provide arbitrary-field Debye formulas together with weak- and strong-field limits that are useful comparisons for the leading magnetic coefficient~\cite{Bandyopadhyay:2017cle}. Lowest-Landau-level analyses of
magnetic-field-induced screening and collective gauge excitations provide a
complementary strong-field limiting picture~\cite{Fukushima:2011nu}, while
strong-field perturbative thermodynamics addresses a distinct parametric limit~\cite{Fraga:2023cef}.

Nonperturbative color screening has also been studied extensively on the
lattice. Gauge-invariant electric and magnetic screening masses can be
extracted from appropriately projected Polyakov-line correlators
~\cite{Maezawa:2010vj,Borsanyi:2015yka}; a broad review of lattice and
weak-coupling descriptions of color screening is given in
Ref.~\cite{Bazavov:2020teh}. These long-distance observables provide useful physical context for the perturbative screening calculation.

In this work $eB/T^2$ is treated as an independent dimensionless parameter: we do not assign the magnetic field any additional power of the QCD coupling $g$. Thus the weak-coupling expansion is performed in $g$ at fixed $T$, $\mu/T$, and $eB/T^2$, while the magnetic-field dependence itself is kept unexpanded. Flavor dependence enters through $q_fB=Q_feB$ inside the Landau-level and proper-time expressions.

The calculation is organized directly around the quadratic operators of the
magnetized electrostatic effective theory. The local $A_0^2$ coefficient is
fixed by the zero-momentum self-energy, while the magnetic field produces two
independent temporal kinetic coefficients, parallel and perpendicular to
$\bm B$, and at finite density an allowed $A_0F_{12}$ mixing operator. The
quark-bearing two-loop contribution is generated as the curvature of the
exchange functional with respect to a constant temporal color source. Its
magnetic part is renormalized by using the corresponding zero-field
ultraviolet subgraphs as local subtraction terms with analytic addbacks,
together with exact SU(3) color projection and the strict dimensional-
regularization treatment of the static exchanged-gluon overlap. We quote
these local matching coefficients separately and denote the angle-dependent
zero of the resulting hard quadratic kernel by $m_E^2(\theta;B)$. At nonzero
$B$ this notation refers to the strict-hard screening pole of the anisotropic
quadratic kernel, while $m_{E,{\rm loc}}^2(B)$ denotes the local $A_0^2$
coefficient. At $B=0$ it reduces to the conventional EQCD mass parameter
$m_E^2(0)$. The numerical analysis is performed for three massless flavors with physical electric charges. The full field scan is evaluated at $\mu/T=0$ and $1$, and
a dedicated density scan extends to $\mu/T=2$ at $eB/T^2=1$ and $4$.

The paper is organized as follows. Section~\ref{sec:screeningDefinition}
defines the screening quantity used throughout, and Sec.~\ref{sec:zeroB}
reviews the zero-field strict-hard result, including its derivation from
the exchange representation. Section~\ref{sec:oneLoopB} collects the
exact one-loop finite-field ingredients, while Sec.~\ref{sec:twoLoopB}
develops the finite-field two-loop exchange contribution and its
renormalization. In Sec.~\ref{sec:assemblyAndTargets} these
ingredients are assembled into the general-angle strict-hard screening pole and
its zero-field reduction is made explicit. The numerical setup and the
resulting phenomenological trends are presented in Secs.~\ref{sec:numerics}
and~\ref{sec:results}, followed by the summary and outlook in\ Sec.~\ref{sec:summaryOutlook}.
\section{Anisotropic hard matching and the static screening pole}
\label{sec:screeningDefinition}
We first define the screening quantity directly in terms of the static gluon
self-energy. The one- and two-loop self-energies are calculated below and
then inserted into this screening condition.

We work with massless QCD in Euclidean signature. The external electromagnetic
field is taken along the $z$ axis, $\bm B=B\hat{\bm z}$, and the quark charge
is $q_f=Q_fe$. Our color normalization is
\begin{align}
 \tr(t^at^b)&=\TR\delta^{ab},
 &\TR&=\frac12,
 \notag\\
 \CA&=N_c,
 &\CF&=\frac{N_c^2-1}{2N_c},
 \notag\\
 \TF&=N_f\TR=\frac{N_f}{2}.
 \label{eq:colorconv}
\end{align}
Fermionic and bosonic Matsubara frequencies are
\begin{equation}
 K_0=(2n+1)\pi T-i\mu_f,\qquad P_0=2\nu\pi T,
\end{equation}
and dimensional regularization is used with $d=3-2\epsilon$ spatial
dimensions and
\begin{equation}
 \int_{\bm p}\equiv
 \left(\frac{e^{\gamma_E}\Lambda^2}{4\pi}\right)^\epsilon
 \int\frac{\dd^d p}{(2\pi)^d}.
 \label{eq:DRmeasure}
\end{equation}
Throughout the analytic calculation, $\Lambda$ denotes the
$\overline{\mathrm{MS}}$ renormalization scale. This notation is kept distinct
from the dimensionless chemical potential $\bar\mu_f\equiv\mu_f/(2\pi T)$.
Later, $\Lambda_q$ will denote the $B$-independent thermal/density scale used
only in the running-coupling illustration.

For a Euclidean four-momentum $K=(K_0,\bm k)$ we denote the magnitude of
its spatial part by
\begin{equation}
 k\equiv|\bm k|,\qquad K^2=K_0^2+k^2.
 \label{eq:momentumConvention}
\end{equation}
The same convention is used for the other loop momenta, e.g.
$p\equiv|\bm p|$ for $P=(P_0,\bm p)$. Bold symbols therefore denote
spatial vectors, while the corresponding italic symbols denote their
magnitudes.

We parameterize the screening momentum by the angle $\theta$ it makes with the magnetic field,
\begin{align}
 \bm k&=k(\sin\theta,0,\cos\theta),
 \qquad k=|\bm k|,
 \notag\\
 k_\parallel&=k\cos\theta,
 \qquad
 k_\perp\equiv|\bm k_\perp|=k\sin\theta,
 \notag\\
 k^2&=k_\parallel^2+k_\perp^2 .
 \label{eq:kangleDefinition}
\end{align}
The static polarization tensor is expanded in loops,
\begin{equation}
 \Pi_{\mu\nu}(0,\bm k;B)
 =\Pi_{\mu\nu}^{(1)}(0,\bm k;B)
 +\Pi_{\mu\nu}^{(2)}(0,\bm k;B)+\cdots .
 \label{eq:selfEnergyLoopExpansion}
\end{equation}
Axial symmetry about $\bm B$ allows two independent coefficients of the
quadratic momentum dependence of the temporal self-energy. We define the
renormalized one-loop coefficients directly by the small-momentum expansion
\begin{align}
 \Pi_{00}^{(1)}(0,\bm k;B)
 ={}&g^2\mathcal A_B
 +g^2z_\parallel(B)k_\parallel^2
 +g^2z_\perp(B)k_\perp^2
 \notag\\
 &+O\!\left(g^2\frac{k^4}{T^2}\right),
 \label{eq:selfEnergyCoefficientsDefinition}
\end{align}
where
\begin{align}
 \mathcal A_B
 &\equiv \frac{1}{g^2}\Pi_{00}^{(1)}(0,\bm0;B),
\notag \\
 z_\parallel(B)
 &\equiv \frac{1}{g^2}
 \left.\frac{\partial\Pi_{00}^{(1)}}{\partial k_\parallel^2}
 \right|_{\bm k=0}^{\overline{\rm MS}},
 \notag\\
 z_\perp(B)
 &\equiv \frac{1}{g^2}
 \left.\frac{\partial\Pi_{00}^{(1)}}{\partial k_\perp^2}
 \right|_{\bm k=0}^{\overline{\rm MS}}.
 \label{eq:oneLoopCoefficientDefinitions}
\end{align}
Thus $\mathcal A_B$ has mass dimension two and is the coefficient of the
zero-momentum $\mathcal{O}(g^2)$ electric self-energy, whereas
$z_\parallel$ and $z_\perp$ are dimensionless renormalized kinetic
coefficients. Their magnetic differences are ultraviolet finite and will be
evaluated explicitly in Sec.~\ref{sec:oneLoopB}.

At finite density and magnetic field the one-loop polarization tensor can
contain an antisymmetric temporal-spatial component. Such a structure has
been exhibited explicitly in the finite-density magnetized polarization
tensor of Wang and Shovkovy~\cite{Wang:2021eud}. In the static limit relevant
here, it mixes $A_0$ with the spatial polarization $A_t$ perpendicular to
both $\bm B$ and the transverse part of $\bm k$. Factoring out the one-loop
QCD coupling, we define
\begin{equation}
 g^2\mathcal M(B)\equiv
 \lim_{k_\perp\to0}\frac{\Pi_{0t}^{(1)}(0,\bm k;B)}{k_\perp}.
 \label{eq:mixSelfEnergyDefinition}
\end{equation}
The corresponding temporal-spatial components are antisymmetric,
\begin{equation}
 \Pi_{t0}^{(1)}=-\Pi_{0t}^{(1)}.
 \label{eq:mixSelfEnergyAntisymmetry}
\end{equation}
Thus $\mathcal M(B)$ has mass dimension one and is the one-loop coefficient
of the linear-in-$k_\perp$ temporal-spatial mixing.

Rather than assigning an angle-dependent quantity to a single EQCD mass
parameter, we organize the hard matching in terms of the local quadratic
operators of the anisotropic electrostatic theory. In the normalization used
throughout this paper their electric part is
\begin{align}
 \mathcal L_{\rm E,hard}^{(2)}={}&
 \frac12 Z_\parallel(B)(D_zA_0)^a(D_zA_0)^a
 \notag\\
 &+\frac12 Z_\perp(B)(D_\alpha A_0)^a(D_\alpha A_0)^a
 \notag\\
 &+\frac12 m_{E,{\rm loc}}^2(B)A_0^aA_0^a
 \notag\\
 &+i\kappa_B A_0^aF_{12}^a+\cdots .
 \label{eq:hardEFTquadratic}
\end{align}
Here $\alpha=1,2$. The ellipsis includes anisotropic purely spatial gauge operators and
higher-dimension terms. They do not affect the electric pole through
$\mathcal{O}(g^4T^2)$ because the spatial entry needed below can be kept at its tree
value at this order. The hard matching coefficients are
\begin{subequations}\label{eq:hardEFTmatching}
\begin{align}
 m_{E,{\rm loc}}^2(B)
 &=g^2\mathcal A_B+g^4c_{B,{\rm strict}}+\mathcal{O}(g^6T^2),
 \label{eq:hardEFTmassLocal}\\
 Z_\parallel(B)&=1+g^2z_\parallel(B)+\mathcal{O}(g^4),
 \label{eq:hardEFTkineticParallel}\\
 Z_\perp(B)&=1+g^2z_\perp(B)+\mathcal{O}(g^4),
 \label{eq:hardEFTkineticPerp}\\
 \kappa_B&=g^2\mathcal M(B)+\mathcal{O}(g^4T).
 \label{eq:hardEFTmixing}
\end{align}
\end{subequations}
Here $Z_\parallel$ and $Z_\perp$ are the full EFT kinetic normalizations,
including their tree-level value $1$; their one-loop corrections are precisely
the $g^2z_\parallel$ and $g^2z_\perp$ terms defined in
Eq.~\eqref{eq:oneLoopCoefficientDefinitions}. The coefficient
$c_{B,{\rm strict}}$ has mass dimension two and denotes the finite
$\overline{\rm MS}$-renormalized $\mathcal{O}(g^4)$ contribution to the local $A_0^2$
operator in the same field normalization. Thus $m_{E,{\rm loc}}^2(B)$ is the
coefficient of the local $A_0^2$ operator, whereas the directional pole also
depends on $Z_\parallel$, $Z_\perp$, and $\kappa_B$. At $B=0$ rotational
invariance gives $Z_\parallel=Z_\perp\equiv Z_0$ and $\kappa_B=0$, so a single
field rescaling exists and the conventional EQCD parameter is
$m_E^2(0)=m_{E,{\rm loc}}^2(0)/Z_0$ order by order.

For a momentum making an angle $\theta$ with $\bm B$ we define
\begin{align}
 Z_\theta(B)={}&Z_\parallel(B)\cos^2\theta
 +Z_\perp(B)\sin^2\theta
 \notag\\
 ={}&1+g^2z_\theta(B)+\mathcal{O}(g^4).
 \label{eq:ZthetaEFT}
\end{align}
Accordingly, the angular one-loop kinetic coefficient is
$z_\theta(B)\equiv z_\parallel(B)\cos^2\theta+z_\perp(B)\sin^2\theta$.
The quadratic one-particle-irreducible kernel in the $(A_0,A_t)$ sector is
therefore
\begin{align}
 \Gamma^{(2)}_{\{0,t\}}&=
 \begin{pmatrix}
 m_{E,{\rm loc}}^2(B)+Z_\theta(B)k^2
 & \kappa_B k\sin\theta\\[1mm]
 -\kappa_B k\sin\theta & k^2+\mathcal{O}(g^2k^2)
 \end{pmatrix}\notag\\
 &+\mathcal{O}(g^6T^2,k^4/T^2).
 \label{eq:selfEnergyMatrixDefinition}
\end{align}
The zero of its electric Schur complement defines the strict-hard screening
pole, which we denote by $k^2=-m_E^2(\theta;B)$. Expanding strictly through
$\mathcal{O}(g^4T^2)$ gives
\begin{align}
 m_E^2(\theta;B)
 ={}&g^2\mathcal A_B
 +g^4\big[c_{B,{\rm strict}}-\mathcal A_Bz_\theta(B)\big]
 \notag\\
 &+g^4\mathcal M^2(B)\sin^2\theta
 +\mathcal{O}(g^6T^2).
 \label{eq:screeningFromSelfEnergy}
\end{align}
The separately resummed soft electrostatic contribution beginning at
$\mathcal{O}(g^3T^2)$ is not included. At $B=0$,
Eq.~\eqref{eq:screeningFromSelfEnergy} reduces to the standard hard EQCD
mass parameter $m_E^2(0)$.

\section{Zero-field screening mass at finite temperature and density}
\label{sec:zeroB}
The zero-field result is useful for two reasons. It fixes the absolute
normalization of the magnetic calculation, and it makes explicit what is meant
here by a \emph{strict-hard} electric mass. We therefore derive the
finite-$T,\mu$ result through $g^4$ before turning on the magnetic field.
The derivation follows the standard strict pole-expansion and finite-density
EQCD matching organization~\cite{Braaten:1995jr,Vuorinen:2003fs}, but we keep the
diagrammatic steps visible because the same organization is used below at
$B\neq0$.
\subsection{One-loop diagrams and the leading electric curvature}
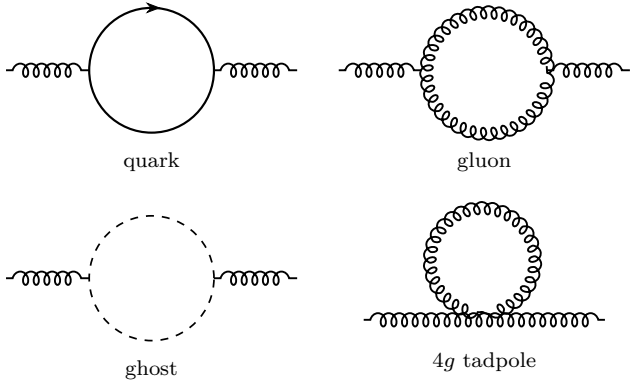
\begin{figure}[h]
	\centering
	\begin{tikzpicture}[scale=1.1]
		\begin{scope}[shift={(0,0)}]
			\draw[gluon] (-1.75,0)--(-.75,0);
			\draw[gluon] (.75,0)--(1.75,0);
			\draw[qline] (0,0) circle[radius=.75];
			\draw[-{Stealth[length=2mm]}] (0.1,.75)--(.11,.75);
			\node[font=\footnotesize] at (0,-1.1) {quark};
		\end{scope}
		
		\begin{scope}[shift={(4.0,0)}]
			\draw[gluon] (-1.75,0)--(-.75,0);
			\draw[gluon] (.75,0)--(1.75,0);
			\draw[thick, gluon] (0.75, .05cm) arc (366.8:-0:0.72cm);
			\node[font=\footnotesize] at (0,-1.1) {gluon};
		\end{scope}
		
		\begin{scope}[shift={(0,-2.5)}]
			\draw[gluon] (-1.75,0)--(-.75,0);
			\draw[gluon] (.75,0)--(1.75,0);
			\draw[ghostline,-{Stealth[length=2mm]}] (0,0) circle[radius=.75];
			\node[font=\footnotesize] at (0,-1.1) {ghost};
		\end{scope}
		
		\begin{scope}[shift={(4.0,-3)}]
			\draw[gluon] (-1.45,0)--(1.45,0);
			\draw[thick, gluon] (0, .09) arc (272:-92:0.62cm);
			\node[font=\footnotesize] at (0,-.5) {$4g$ tadpole};
		\end{scope}
	\end{tikzpicture}
	\caption{One-loop gluon self-energy topologies: quark, gluon, and ghost loops, and the four-gluon tadpole.}
	\label{fig:oneLoopZero}
\end{figure}

The one-loop temporal gluon self-energy contains the quark bubble, gluon
bubble, ghost loop and four-gluon tadpole shown schematically in
Fig.~\ref{fig:oneLoopZero}. Only the quark loop carries the chemical
potential.
For one massless flavor the quark bubble is
\begin{align}
 \Pi_{00}^{q,ab}(P)
 &=-g^2\TR\delta^{ab}\notag\\&\times\sumint_{\{K\}}
 \tr_D\!\left[\gamma_0S(K)\gamma_0S(K-P)\right],
\end{align}
with
\begin{align}
 S(K)&=\frac{-i\slashed K}{K^2},\qquad
 K_0=(2n+1)\pi T-i\mu_f .
 \label{eq:zeroBquarkbubble}
\end{align}
Here and below the unbraced and braced sum-integrals denote bosonic and
fermionic Matsubara sums, respectively,
\begin{equation*}
 \sumint_K \equiv T\sum_{K_0=2\pi nT}\int_{\bm k},
 \qquad
 \sumint_{\{K\}} \equiv
 T\sum_{K_0=(2n+1)\pi T-i\mu_f}\int_{\bm k}.
\end{equation*}
Thus braces on $\{K\}$ indicate a fermionic sum-integral (including the
chemical-potential shift), whereas an unbraced $K$ denotes a bosonic
sum-integral. The spatial measure $\int_{\bm k}$ is the
$\overline{\mathrm{MS}}$ measure defined in Eq.~\eqref{eq:DRmeasure}.
In the static limit $P=(0,\bm p)$, followed by $\bm p\to0$, we define
 \begin{align}
 I_n&\equiv\sumint_K\frac1{(K^2)^n},
 &\quad \widetilde I_n(\mu_f)&\equiv\sumint_{\{K\}}\frac1{(K^2)^n},\notag\\
 J_n&\equiv\sumint_K\frac{K_0^2}{(K^2)^{n+1}},
 &\quad \widetilde J_n(\mu_f)&\equiv\sumint_{\{K\}}\frac{K_0^2}{(K^2)^{n+1}},\notag\\
& \widetilde N_n(\mu_f)\equiv
 \sumint_{\{K\}}\frac{K_0}{(K^2)^n}.
 \label{eq:zeroBsumintegrals}
 \end{align}
The last sum-integral is needed at finite density because differentiating a fermion
propagator with respect to its temporal momentum produces numerators linear in
$K_0$. It vanishes at $\mu_f=0$, but need not vanish once the fermionic
frequency is shifted by the chemical potential.
The Yang-Mills and quark pieces reduce to
\begin{align}
 \Pi_{00}^{\rm YM}(0,\bm0)
 &=g^2\CA(D-2)(I_1-2J_1),\notag\\
 \Pi_{00}^{q_f}(0,\bm0)
 &=-4g^2\TR\,[\widetilde I_1(\mu_f)-2\widetilde J_1(\mu_f)].
 \label{eq:zeroB1loopsumintegrals}
\end{align}
with $D=4-2\epsilon$. Spatial integration by parts gives
\begin{equation}
 J_n=\left(1-\frac{d}{2n}\right)I_n,\qquad
 \widetilde J_n=\left(1-\frac{d}{2n}\right)\widetilde I_n,
 \label{eq:zeroBIBP}
\end{equation}
with $d=3-2\epsilon$. At $\epsilon=0$,
\begin{equation}
 I_1=\frac{T^2}{12},\qquad
 \widetilde I_1(\mu_f)
 =-\frac{T^2}{24}-\frac{\mu_f^2}{8\pi^2},
 \label{eq:zeroBtadpoles}
\end{equation}
which gives
\begin{align}
 \Pi_{00}^{(1)}(0,\bm0)
 &=\Pi_{00}^{\rm YM}(0,\bm0)
 +\sum_f\Pi_{00}^{q_f}(0,\bm0)\notag\\
 &=g^2\mathcal A_0.
 \label{eq:zeroBOneLoopPi00}
\end{align}
The corresponding zero-field one-loop coefficient is
\begin{equation}
 \mathcal A_0
 =\frac{\CA T^2}{3}+\TR\sum_f
 \left(\frac{T^2}{3}+\frac{\mu_f^2}{\pi^2}\right).
 \label{eq:A0}
\end{equation}
\subsection{Two-loop diagrams and strict pole expansion}
\begin{figure*}[t]
\centering
  \resizebox{0.7\textwidth}{!}{
\begin{tikzpicture}[x=1.3cm,y=1.3cm]
\begin{scope}[xshift=-4.7cm,yshift=1.25cm]
  \coordinate (T) at (0,.78); \coordinate (B) at (0,-.78);
  \coordinate (L) at (-.78,0); \coordinate (R) at (.78,0);
  \draw[gluon] (-1.55,0)--(L);
  \draw[gluon] (R)--(1.55,0);
  \draw[qline] (T)--(B);
  \draw[-{Stealth[length=2mm,width=2mm]},line width=.72pt] (0,.42)--(0,.06);
  \draw[-{Stealth[length=2mm,width=2mm]},line width=.72pt] (.65,.4) --(.6,.5);
  \draw[thick, gluon] (0., -.97cm) arc (270:90:.95cm);
  \draw[thick] (0cm, 1cm) arc (270:90:-1cm);
  \node[font=\small] at (0,-1.22) {(a) mixed $qg$};
\end{scope}
\begin{scope}[xshift=0cm,yshift=1.25cm]
  \coordinate (T) at (0,.82); \coordinate (B) at (0,-.82);
  \coordinate (L) at (-.82,0); \coordinate (R) at (.82,0);
  \draw[gluon] (-1.7,0)--(L);
  \draw[gluon] (R)--(1.7,0);
  \draw[gluon] (T)--(B);
  \draw[-{Stealth[length=2mm,width=2mm]},line width=.72pt] (.6,.53)--(.7,.43);
  \draw[-{Stealth[length=2mm,width=2mm]},line width=.72pt] (-.6,-.53)--(-.7,-.43);
  \draw[qline] (0,0) circle[radius=.8];
  \node[font=\small] at (0,-1.22) {(b) QED-like vertex};
\end{scope}
\begin{scope}[xshift=4.7cm,yshift=1.25cm]
  \coordinate (L) at (-.82,0); \coordinate (R) at (.82,0);
  \coordinate (U1) at (-.6,.5); \coordinate (U2) at (.6,.5);
  \draw[gluon] (-1.7,0)--(L);
  \draw[gluon] (R)--(1.7,0);
  \draw[gluon] (U1) .. controls (0,.1) .. (U2);
  \draw[-{Stealth[length=2mm,width=2mm]},line width=.72pt] (-.1,.8)--(.1,.8);
  \draw[-{Stealth[length=2mm,width=2mm]},line width=.72pt] (.1,-.8)--(-.1,-.8);
  \draw[qline] (0,0) circle[radius=.8];
  \node[font=\small] at (0,-1.22) {(c) QED-like self-energy};
\end{scope}
\begin{scope}[xshift=-2.5cm,yshift=-2.5cm]
  \coordinate (V)  at (0,-.72);
  \coordinate (QC) at (0,.78);
  \coordinate (QL) at (-.22,.78);
  \coordinate (QR) at (.22,.78);

  \draw[gluon] (-1.65,-.72)--(V);
  \draw[gluon] (V)--(1.65,-.72);

  \draw[gluon] (V) .. controls (-.88,-.42) and (-.88,.42) .. (-.4,.8);
  \draw[gluon] (.4,.8) .. controls (.88,.42) and (.88,-.42) .. (0,-.72);

  \draw[qline] (QC) circle[radius=.4];
  \draw[-{Stealth[length=2mm,width=2mm]},line width=.72pt]
  (-.07,1.18) -- (.08,1.18);
  \draw[-{Stealth[length=2mm,width=2mm]},line width=.72pt]
  (.07,.39) -- (-.075,.39);

  \node[font=\small] at (0,-1.18) {(d) $4g$ contact};
\end{scope}
\begin{scope}[xshift=2.5cm,yshift=-2.5cm]
  \coordinate (L)  at (-.76,0);
  \coordinate (R)  at (.76,0);
  \coordinate (QC) at (0,.84);
  \coordinate (QL) at (-.22,.84);
  \coordinate (QR) at (.22,.84);

  \draw[gluon] (-1.55,0)--(L);
  \draw[gluon] (R)--(1.55,0);

  \draw[gluon] (L) .. controls (-.84,.57) and (-.44,.75) .. (-.38,.88);
  \draw[gluon] (.38,.88) .. controls (.44,.75) and (.84,.57) .. (R);

  \draw[gluon] (R) .. controls (.72,-.86) and (-.72,-.86) .. (L);

  \draw[qline] (QC) circle[radius=.4];
  \draw[-{Stealth[length=2mm,width=2mm]},line width=.72pt]
  (-.07,1.23) --(.075,1.23);
  \draw[-{Stealth[length=2mm,width=2mm]},line width=.72pt]
  (.07,.45)--(-.075,.45);

  \node[font=\small] at (0,-1.18) {(e) $3g$-$3g$};
\end{scope}
\end{tikzpicture}}
\caption{Two-loop gluon-self-energy topologies containing quark lines.}
\label{fig:twoLoopZero}
\end{figure*}
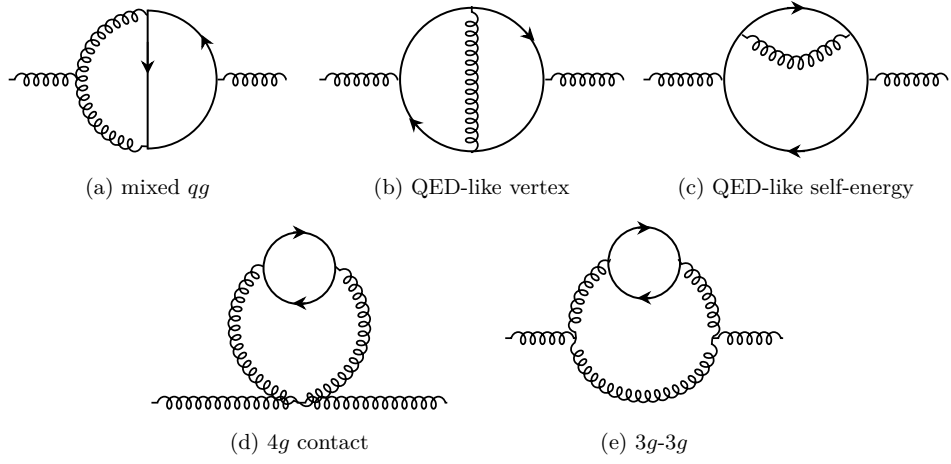

The five quark-bearing two-loop self-energy topologies are displayed in
Fig.~\ref{fig:twoLoopZero}. Two are QED-like and three are genuinely
non-Abelian. At $B=0$ they must be combined with the purely bosonic two-loop
graphs and the coupling counterterm. Specializing the strict condition
Eq.~\eqref{eq:screeningFromSelfEnergy} to $B=0$ gives the standard EQCD
matching expansion used below.

The one-loop momentum derivatives defining $z_\parallel$ and $z_\perp$ in
Eq.~\eqref{eq:oneLoopCoefficientDefinitions} coincide at $B=0$; their common
zero-field value follows from the same one-loop graphs. With
\begin{align}
 \bar\mu_f&\equiv\frac{\mu_f}{2\pi T},
 &z_f&\equiv\frac12-i\bar\mu_f,
 \notag\\
 L&\equiv\ln\frac{\Lambda}{4\pi T}.
 \label{eq:zeroBshort}
\end{align}
the quark and Yang-Mills pieces are
\begin{align}
 \left.\partial_{k^2}\Pi_{00}^{(1),q}\right|_{\bm0}
 =\frac{g^2\TR}{(4\pi)^2}\frac43
 \left[\frac1\epsilon+2L-1
 -\psi(z_f)-\psi(z_f^*)\right],\hspace{1cm}\notag\\
 \left.\partial_{k^2}\Pi_{00}^{(1),g+\mathrm{gh}}\right|_{\bm0}
 =\frac{g^2\CA}{(4\pi)^2}
 \left[-\frac{5}{3\epsilon}+\frac13-\frac{10}{3}(L+\gamma_E)\right].\hspace{1cm}
 \label{eq:zeroBwave}
\end{align}

\subsection{Two-loop exchange curvature and ordinary Feynman-gauge self-energy}
\label{sec:zeroBUVforest}
The quark-bearing two-loop self-energy diagrams in
Fig.~\ref{fig:twoLoopZero} are generated from a single two-loop vacuum
skeleton. This is the organization that will also be used at nonzero magnetic
field. We introduce a constant diagonal temporal color source
\begin{equation}
 A_0=a\mathsf H,\qquad \varphi\equiv ga,
 \label{eq:zeroBbackground}
\end{equation}
where $g$ is the QCD gauge coupling, $a$ is the constant source amplitude,
and $\mathsf H$ is a traceless Hermitian Cartan generator of SU(3). We
choose the normalized direction
\begin{equation}
 \mathsf H=t^3=\mathrm{diag}\!\left(\frac12,-\frac12,0\right),
 \qquad \tr\mathsf H^2=\frac12,
 \label{eq:sourceGeneratorH}
\end{equation}
which entails no loss of generality at vanishing source by global color
symmetry. The eigenvalues $h_i$ of $\mathsf H$ specify the shifts
$\varphi h_i$ of the Euclidean energies of quarks with color $i$. For the two quark
lines in the exchange graph, carrying colors $i$ and $j$, we define
\begin{align}
 P_i&=(P_0+\varphi h_i,\bm p),
 \notag\\
 Q_j&=(Q_0+\varphi h_j,\bm q),
 \notag\\
 R_{ij}&=P_i-Q_j,
 \label{eq:zeroBsourceMomenta}
\end{align}
with $P_0=(2n+1)\pi T-i\mu$ and $Q_0=(2m+1)\pi T-i\mu$ for a common quark
chemical potential $\mu$.

\begin{figure}[htbp]
\centering
\begin{tikzpicture}[scale=1.08]
  \draw[qline] (0,0) circle[radius=1.05];
  \coordinate (vL) at (-.78,.70);
  \coordinate (vR) at (.78,.70);
  \draw[gluon] (vL) .. controls (-.38,.08) and (.38,.08) .. (vR);
  \fill (vL) circle[radius=1.2pt];
  \fill (vR) circle[radius=1.2pt];
  \draw[-{Stealth[length=2.0mm,width=1.5mm]},line width=.72pt]
       (-.82,.66) arc[start angle=141,end angle=102,radius=1.05];
  \draw[-{Stealth[length=2.0mm,width=1.5mm]},line width=.72pt]
       (.82,-.66) arc[start angle=-39,end angle=-78,radius=1.05];
  \node[font=\small] at (-1.15,-.48) {$P$};
  \node[font=\small] at (1.15,-.48) {$Q$};
  \node[font=\small,fill=white,inner sep=1.5pt] at (0,-.10) {$R$};
  \node[font=\small] at (0,-1.42) {$\mathscr G_{\rm exch}^{(0)}(\varphi)$};
\end{tikzpicture}
\caption{Zero-field two-loop quark-gluon exchange skeleton used in the
temporal-source construction.}
\label{fig:exchangeVacuumSkeleton}
\end{figure}
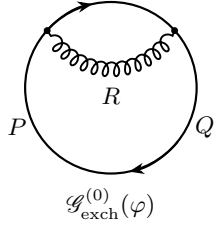

The shifted-frequency exchange functional is most naturally interpreted in
background-field Feynman gauge for the gluon. The constant temporal
field is used only as a source and is set to zero after differentiation. The
ordinary covariant-Feynman self-energy is the quantity entering the strict
screening pole; the finite local conversion between the background-field and
ordinary organizations is derived explicitly below for the magnetic
difference. At $B=0$ the ordinary Feynman-gauge coefficient is fixed
independently by the direct sum-integral calculation in
Sec.~\ref{sec:zeroB}. The auxiliary zero-field exchange functional is
\begin{align}
 \mathscr G_{\rm exch}^{(0)}(\varphi)
 &=\frac{g^2}{2}\sum_f\sum_{i,j}\mathcal C_{ij}
 \sumint_{\{P\}}\sumint_{\{Q\}}\notag\\
 &\quad\times D_{\mu\nu,0}(R_{ij})\,
 \tr_D\!\left[\gamma_\mu S_{f,0}(P_i)
 \gamma_\nu S_{f,0}(Q_j)\right]\notag\\
 &=\frac{g^2}{2}\sum_f\sum_{i,j}\mathcal C_{ij}
 \sumint_{\{P\}}\sumint_{\{Q\}}
 \mathcal F_0(P_i,Q_j).
 \label{eq:zeroBexchangeFunctional}
\end{align}
Here $S_{f,0}(P_i)=-i\slashed P_i/P_i^2$ denotes the free quark
propagator with its source-shifted energy, and
$D_{\mu\nu,0}(R_{ij})=\delta_{\mu\nu}/R_{ij}^2$ is the exchanged-gluon
propagator in background-field Feynman gauge. The factor $1/2$ is the
exchange symmetry factor, arising from the second-order expansion of
the quark-gluon interaction. Contracting the propagators and taking the
Dirac trace gives the integrand
\begin{equation}
 \mathcal F_0(P_i,Q_j)
 =4(D-2)\frac{P_i\!\cdot Q_j}
 {P_i^2Q_j^2R_{ij}^2}.
 \label{eq:zeroBexchangeIntegrand}
\end{equation}
The corresponding exchanged-gluon color weight is
\begin{equation}
 \mathcal C_{ij}=\sum_a(t^a)_{ij}(t^a)_{ji}.
\end{equation}
The superscript $(0)$ denotes the zero-magnetic-field limit of the
same source-dependent exchange functional,
$\mathscr G_{\rm exch}^{(0)}(\varphi)
\equiv\mathscr G_{\rm exch}(\varphi,0)$. The exchange skeleton is
shown in Fig.~\ref{fig:exchangeVacuumSkeleton}, with fermionic momenta
$P$ and $Q$ and exchanged-gluon momentum $R=P-Q$.

The second derivative of the source-shifted exchange functional defines the
background-field Feynman-gauge exchange curvature. It is convenient to write
\begin{equation}
 \mathcal I_0\equiv
 \sum_f\sum_{i,j}\mathcal C_{ij}
 \sumint_{\{P\}}\sumint_{\{Q\}}
 \left.\partial_\varphi^2\mathcal F_0(P_i,Q_j)\right|_{\varphi=0}.
 \label{eq:zeroBexchangeCurvatureIntegral}
\end{equation}
At fixed loop variables, the source curvature $K_f(h_i,h_j)$ of either
the zero-field or magnetized exchange kernel is a homogeneous quadratic
function of the two color eigenvalues. With $K_{ab,f}\equiv K_f(a,b)$,
the SU(3) color projection is
\begin{equation}
 \sum_{ij}\mathcal C_{ij}K_f(h_i,h_j)
 =\frac34(K_{10,f}+K_{01,f})-\frac1{12}K_{11,f}.
 \label{eq:colorprojectionpub}
\end{equation}
This purely color-algebraic relation will also be used at finite $B$.
In terms of $\mathcal I_0$, differentiating the exchange functional
with respect to the shifted-frequency source $\varphi$ gives
\begin{align}
 \Pi_{00,\rm BF,bare}^{(2),q}(0,\bm0;\mu)
 &=-g^2\left.\frac{\partial^2\mathscr G_{\rm exch}^{(0)}(\varphi)}
 {\partial\varphi^2}\right|_{\varphi=0}\notag\\
 &=-\frac{g^4}{2}\,\mathcal I_0.
 \label{eq:zeroBPi2FromI0}
\end{align}
Here the subscript $\rm BF$ denotes background-field Feynman gauge. The
quantity in Eq.~\eqref{eq:zeroBPi2FromI0} is an auxiliary background-field
curvature, not yet the ordinary covariant-Feynman self-energy. The exchange
graph carries $g^2$, while the conversion from the physical source amplitude
$a$ to $\varphi=ga$ contributes $\partial_a^2=g^2\partial_\varphi^2$.
The local background-to-ordinary conversion needed to obtain the
Feynman-gauge self-energy entering the strict pole is evaluated explicitly
below at zero field.

After the background-to-ordinary conversion, the ultraviolet pole cancels in the assembled strict pole through the usual QCD coupling renormalization and the one-loop momentum-derivative contribution. No separate counterterm functional is required for the exchange representation.

To evaluate the curvature we write
\begin{align*}
 \mathscr H_0(P,Q)&\equiv
 4(D-2)\frac{P\!\cdot Q}{P^2Q^2},\qquad R=P-Q,\\
 \mathcal F_0&=\frac{\mathscr H_0}{R^2}.
\end{align*}
and define
\begin{equation*}
 {\cal Q}_{ij}\equiv h_i\partial_{P_0}+h_j\partial_{Q_0},
 \qquad \Delta h_{ij}=h_i-h_j.
\end{equation*}
During the source differentiation $P_0$, $Q_0$, and $R_0$ are treated as
independent kernel arguments, with $R_0=P_0-Q_0$ imposed afterwards. One then
finds
\begin{align}
 \left.\partial_\varphi^2\mathcal F_0\right|_{\varphi=0}
 ={}&
 \frac{{\cal Q}_{ij}^{\,2}\mathscr H_0}{R^2}
 -\frac{4\Delta h_{ij}R_0}{R^4}\,
 {\cal Q}_{ij}\mathscr H_0
 \notag\\
 &+(\Delta h_{ij})^2\mathscr H_0
 \left(
 \frac{8R_0^2}{R^6}-\frac{2}{R^4}
 \right).
 \label{eq:zeroBsourceDerivative}
\end{align}
In the background-field source organization of Eq.~\eqref{eq:zeroBsourceDerivative}, the first term describes two
insertions on the quark sector, corresponding to the QED-like vertex and
self-energy topologies [Figs.~\ref{fig:twoLoopZero}(b) and (c)].
The second term has one insertion on the quark sector and one on the
exchanged gluon [Fig.~\ref{fig:twoLoopZero}(a)], while the third has
two insertions on the exchanged gluon, associated with the four-gluon
contact and two-three-gluon-vertex topologies [Figs.~\ref{fig:twoLoopZero}(d)
and (e)].

Using the SU(3) color sums collected in
Appendix~\ref{app:zeroBalgebra}, the curvature $\mathcal I_0$ defined in
Eq.~\eqref{eq:zeroBexchangeCurvatureIntegral} reduces to
{\allowdisplaybreaks[4]
\begin{align}
 {\cal I}_0
 =\sumint_{\{P\}}\sumint_{\{Q\}}
 \Bigg\{&
 \frac{\CF\TF}{R^2}
 (\partial_{P_0}^2+\partial_{Q_0}^2)\mathscr H_0
 \notag\\
 &-\frac{\TF(\CA-2\CF)}{R^2}
 \partial_{P_0}\partial_{Q_0}\mathscr H_0
 \notag\\
 &-\frac{2\CA\TF R_0}{R^4}
 (\partial_{P_0}-\partial_{Q_0})\mathscr H_0
 \notag\\
 &+\CA\TF\mathscr H_0
 \left(\frac{8R_0^2}{R^6}-\frac{2}{R^4}\right)
 \Bigg\}.
 \label{eq:zeroBI0reduced}
\end{align}}
The explicit temporal derivatives are given in
Appendix~\ref{app:zeroBalgebra}.

We now reduce Eq.~\eqref{eq:zeroBI0reduced} to scalar thermal
sum-integrals and include the zero-field form of the local
background-to-ordinary conversion before identifying the result with the
ordinary Feynman-gauge self-energy. The sum-integrals are defined in
Eq.~\eqref{eq:zeroBsumintegrals}. For a common chemical potential we use
\begin{equation}
 \bar\mu\equiv\frac{\mu}{2\pi T},\qquad
 z_\mu\equiv\frac12-i\bar\mu.
 \label{eq:zeroBmuDefs}
\end{equation}
The required sum-integrals are given in Appendix~\ref{app:zeroBalgebra}.
Using $2P\!\cdot Q=P^2+Q^2-R^2$, shift invariance of the double
sum-integral, $P\leftrightarrow Q$ symmetry, and the spatial
integration-by-parts identity Eq.~\eqref{eq:zeroBIBP}, the
\emph{background-field Feynman-gauge curvature} reduces to
 \begin{align}
 \Pi_{00,\mathrm{BF,bare}}^{(2),q}
 &=g^4\bigg\{
 8(1-\epsilon)\CA\TF\,\widetilde I_1
 \big(I_2-4J_2\big)\notag\\
 &\quad+8(1-\epsilon)\CF\TF
 (I_1-\widetilde I_1)(\widetilde I_2-4\widetilde J_2)\notag\\
 &\quad-8(1-\epsilon)\TF(\CA-2\CF)
 [\widetilde N_2(\mu)]^2\bigg\}.
 \label{eq:zeroBbgPi2sumintegrals}
  \end{align}
This is the curvature generated by the shifted-frequency exchange functional;
it is not yet the ordinary covariant-Feynman self-energy. In particular,
Eq.~\eqref{eq:zeroBIBP} gives
$I_2-4J_2=-2\epsilon I_2$ and
$\widetilde I_2-4\widetilde J_2=-2\epsilon\widetilde I_2$.
The corresponding terms are finite after the sum-integrals are evaluated,
even though they appear evanescent before the ultraviolet poles of $I_2$ and
$\widetilde I_2$ are exposed.

The ordinary Feynman-gauge three-gluon source vertex also contains the
longitudinal pieces $-\delta_{0\alpha}R_\beta-\delta_{0\beta}R_\alpha$,
absent from the shifted background-field propagator. When the mixed
quark-gluon and two-gluon source-insertion sectors are combined, these terms
leave a local correction proportional to the one-loop quark self-energy.
Using Eq.~\eqref{eq:zeroB1loopsumintegrals} for the latter, the zero-field
conversion reads
\begin{align}
 \Pi_{00,\mathrm{bare}}^{(2),q}
 &=\Pi_{00,\mathrm{BF,bare}}^{(2),q}
 +2\CA g^2 I_2\,\Pi_{00}^{(1),q}\notag\\
 &=\Pi_{00,\mathrm{BF,bare}}^{(2),q}
 -8\CA\TF g^4 I_2
 (\widetilde I_1-2\widetilde J_1).
 \label{eq:zeroBbgToFeynman}
\end{align}
Substituting Eq.~\eqref{eq:zeroBbgPi2sumintegrals} into
Eq.~\eqref{eq:zeroBbgToFeynman} and collecting terms gives the ordinary
Feynman-gauge quark-bearing two-loop self-energy,
\begin{align}
 \Pi_{00,\mathrm{bare}}^{(2),q}
 &=g^4\bigg\{
 8\CA\TF\left[2I_2\widetilde J_1
 -\epsilon\widetilde I_1I_2
 -4(1-\epsilon)\widetilde I_1J_2\right]\notag\\
 &+8(1-\epsilon)\CF\TF
 (I_1-\widetilde I_1)(\widetilde I_2-4\widetilde J_2)\notag\\
 &-8(1-\epsilon)\TF(\CA-2\CF)
 [\widetilde N_2(\mu)]^2\bigg\}.
 \label{eq:zeroBquarkPi2sumintegrals}
\end{align}
Equation~\eqref{eq:zeroBbgToFeynman} is the explicit $B=0$ validation of the
conversion used below at finite field. The ultraviolet pole of the ordinary
self-energy is generated by the local conversion term, whereas the
background-field curvature in Eq.~\eqref{eq:zeroBbgPi2sumintegrals} has a
finite $\epsilon\to0$ limit. The apparently evanescent thermal pieces in
these equations are also the zero-field prototype of the sum-first
three-dimensional contribution discussed in
Sec.~\ref{sec:finiteBsubtraction}.

The term proportional to $[\widetilde N_2]^2$ arises from the same
temporal-momentum derivatives listed in Appendix~\ref{app:zeroBalgebra}.
It vanishes at $\mu=0$, whereas at finite density the shift of the fermionic
Matsubara frequencies allows this sum-integral to be nonzero.

Substitution of these sum-integrals gives the quark-bearing contribution in closed
form,
\begin{align}
 \frac{(4\pi)^2\Pi_{00,\mathrm{bare}}^{(2),q}}
 {g^4T^2}
 &={}
 \frac{2\CA\TF(1+12\bar\mu^2)}{3\epsilon}
 \notag\\
 &\hspace{-1cm}+\frac{\CA\TF}{9}\Bigg\{
 12(1+12\bar\mu^2)(2L+\gamma_E) +18(1+12\bar\mu^2)
 \notag\\
 &
-6
 +144\big[\zeta'(-1,z_\mu)
 +\zeta'(-1,z_\mu^*)\big]
 \Bigg\}
 \notag\\
 &-2\CF\TF(1+12\bar\mu^2).
 \label{eq:zeroBquarkPi2closed}
\end{align}
Since the aim of the present work is to determine the magnetic-field dependence of the strict-hard contribution to the screening mass, we restrict the detailed two-loop discussion to diagrams containing quark lines. Purely gluonic and ghost diagrams do not couple directly to the external electromagnetic field and therefore do not generate any additional $B$ dependence. Their known zero-field contribution can be taken from Refs.~\cite{Braaten:1995jr,Vuorinen:2003fs} and is
\begin{align}
 \frac{(4\pi)^2\Pi_{00,\mathrm{bare}}^{(2),g}}
 {g^4T^2}
 ={}&\frac{2\CA^2}{3\epsilon}
 +\frac{2\CA^2}{3}
 \left[1+2\gamma_E+4L\right.
 \notag\\
 &\hspace{25mm}\left.+2\frac{\zeta'(-1)}{\zeta(-1)}\right].
 \label{eq:zeroBgluePi2closed}
\end{align}
Adding Eqs.~\eqref{eq:zeroBquarkPi2closed} and~\eqref{eq:zeroBgluePi2closed} yields the full bare two-loop temporal
self-energy,
\begin{widetext}
\begin{align}
 \frac{(4\pi)^2\Pi_{00,\mathrm{bare}}^{(2)}(0,\bm0;\mu)}{g^4T^2}
 ={}&\frac{2\CA[\CA+\TF(1+12\bar\mu^2)]}{3\epsilon}+\frac{2\CA^2}{3}
 \left[1+2\gamma_E
 +2\frac{\zeta'(-1)}{\zeta(-1)}+4L\right]\notag\\
 &+\frac{\CA\TF}{9}\Bigg\{
 12(1+12\bar\mu^2)(2L+\gamma_E)
 +18(1+12\bar\mu^2)-6\notag\\
 &\qquad+144\big[\zeta'(-1,z_\mu)+\zeta'(-1,z_\mu^*)\big]\Bigg\}
 -2\CF\TF(1+12\bar\mu^2) .
 \label{eq:zeroBbarePi2}
\end{align}
\end{widetext}
At $\mu=0$, Eq.~\eqref{eq:zeroBbarePi2} agrees with the Braaten-Nieto
two-loop self-energy calculation~\cite{Braaten:1995jr}. After the one-loop
pole-derivative term and coupling renormalization are included in the strict
pole combination, the finite-density result agrees with Vuorinen's EQCD mass
matching coefficient~\cite{Vuorinen:2003fs}. This provides the zero-field
validation of the exchange-curvature organization used below at finite
magnetic field.

At zero field the calculation can be completed analytically in dimensional
regularization, so no separate integrand-level subtraction is needed. The
$1/\epsilon$ pole in Eq.~\eqref{eq:zeroBbarePi2} combines with the one-loop
momentum-derivative term and the coupling counterterm in the strict pole
condition. The following subsection displays this cancellation explicitly.
\subsection{Pole cancellation and the finite zero-field coefficient}
For the ultraviolet cancellation it is useful to collect the one- and two-loop
quantities in the form
\begin{subequations}
\label{eq:zeroBpoleIngredients}
\begin{align}
 \Pi_{00}^{(1)}(0,\bm0)&=g^2T^2(M_0+\epsilon M_1),
 \label{eq:zeroBpoleIngredientsA}\\
 \left.\partial_{k^2}\Pi_{00}^{(1)}\right|_{\bm0}
 &=\frac{g^2}{(4\pi)^2}\left(\frac{W_{-1}}\epsilon+W_0\right),
 \label{eq:zeroBpoleIngredientsB}\\
 \Pi_{00,\mathrm{bare}}^{(2)}(0,\bm0)
 &=\frac{g^4T^2}{(4\pi)^2}\left(\frac{B_{-1}}\epsilon+B_0\right).
 \label{eq:zeroBpoleIngredientsC}
\end{align}
\end{subequations}
The coefficients appearing in Eqs.~\eqref{eq:zeroBpoleIngredientsA}-\eqref{eq:zeroBpoleIngredientsC} are
\begin{subequations}
\begin{align}
 M_0&=\frac{\CA+\TF(1+12\bar\mu^2)}{3},
 \\
 M_1&=\frac13\Bigg\{
 2\CA\left(L+\frac{\zeta'(-1)}{\zeta(-1)}\right)
 \notag\\
 &\qquad+\TF(1+12\bar\mu^2)(2L+1)
 \notag\\
 &\qquad+24\TF
 \big[\zeta'(-1,z_\mu)+\zeta'(-1,z_\mu^*)\big]
 \Bigg\},\\
 W_{-1}&=\frac{-5\CA+4\TF}{3},\\
 W_0&=\frac{\CA}{3}[1-10(L+\gamma_E)]
 \notag\\
 &\qquad+\frac{4\TF}{3}
 \big[2L-1-\psi(z_\mu)-\psi(z_\mu^*)\big],
 \\
 B_{-1}&=
 \frac{2\CA[\CA+\TF(1+12\bar\mu^2)]}{3},\\
 B_0&=\frac{2\CA^2}{3}
 \left[
 1+2\gamma_E
 +2\frac{\zeta'(-1)}{\zeta(-1)}
 +4L
 \right]\notag\\
 &\quad+\frac{\CA\TF}{9}\Bigg\{
 12(1+12\bar\mu^2)(2L+\gamma_E)
 \notag\\
 &\hspace{12mm}
 +18(1+12\bar\mu^2)-6
 \notag\\
 &\hspace{12mm}
 +144[\zeta'(-1,z_\mu)+\zeta'(-1,z_\mu^*)]
 \Bigg\}\notag\\
 &\quad-2\CF\TF(1+12\bar\mu^2).
 \label{eq:zeroBMZ}
\end{align}
\end{subequations}
Here $W_{-1}$ and $W_0$ denote the pole and finite Laurent coefficients of
the zero-field one-loop momentum derivative in the ordinary covariant
Feynman gauge used for the self-energy entering the strict pole. Consequently
$W_{-1}=(-5\CA+4\TF)/3$, rather than the background-Feynman-gauge value
$-\beta_0$.
The bare and renormalized couplings are related by
\begin{equation}
 g_0^2=g^2\left[1-\frac{\beta_0g^2}{(4\pi)^2\epsilon}\right],
 \qquad \beta_0=\frac{11\CA-4\TF}{3}.
 \label{eq:zeroBgct}
\end{equation}
Substituting Eqs.~\eqref{eq:zeroBpoleIngredientsA}-\eqref{eq:zeroBpoleIngredientsC} into the strict pole
condition, Eq.~\eqref{eq:screeningFromSelfEnergy}, and expressing the bare coupling
through Eq.~\eqref{eq:zeroBgct}, the coefficient of the $1/\epsilon$ pole at
order $g^4$ is
$B_{-1}-M_0(\beta_0+W_{-1})$. Here $B_{-1}$ comes from the unrenormalized
two-loop self-energy, $\beta_0M_0$ from coupling renormalization of the
leading one-loop term, and $M_0W_{-1}$ from the one-loop momentum-derivative
term. The pole therefore cancels algebraically,
\begin{align}
 B_{-1}&-M_0(\beta_0+W_{-1})\notag\\
 ={}&
 \frac{2\CA[\CA+\TF(1+12\bar\mu^2)]}{3}
 -2\CA M_0
 =0.
 \label{eq:zeroBpolecancel}
\end{align}
Using the zero-field strict pole expansion, Eq.~\eqref{eq:screeningFromSelfEnergy},
in the standard form of Ref.~\cite{Braaten:1995jr}, the finite result is
\begin{align}
 m_E^2
 ={}&\left[\Pi_{00}^{(1)}(0,\bm0)+\Pi_{00}^{(2)}(0,\bm0)
 -\Pi_{00}^{(1)}(0,\bm0)
 \left.\partial_{k^2}\Pi_{00}^{(1)}\right|_{\bm0}\right]_{\overline{\rm MS}}
 \notag\\
 ={}&g^2T^2M_0
 +\frac{g^4T^2}{(4\pi)^2}
 \big[B_0-(\beta_0+W_{-1})M_1-M_0W_0\big]
 \notag\\
 &+\mathcal{O}(g^6T^2).
 \label{eq:zeroBassembly}
\end{align}
The $\mathcal{O}(\epsilon)$ part of the one-loop result is essential because it multiplies
the ultraviolet poles. For later use, it is helpful to split the finite
$\mathcal{O}(g^4)$ coefficient in the same $\overline{\rm MS}$ convention used for the
finite-field kernel. We define
\begin{equation}
\begin{aligned}
 z_0&\equiv\frac{W_0}{(4\pi)^2},\\
 \frac{c_{0,\rm strict}}{T^2}
 &\equiv\frac{1}{(4\pi)^2}
 \big[B_0-(\beta_0+W_{-1})M_1\big].
\end{aligned}
\label{eq:zeroBfiniteSplit}
\end{equation}
Then Eq.~\eqref{eq:zeroBassembly} is equivalently
\begin{equation}
 m_E^2=g^2T^2M_0+g^4\big[c_{0,\rm strict}-T^2M_0z_0\big]
 +\mathcal{O}(g^6T^2).
\label{eq:zeroBsplitPole}
\end{equation}

For independent flavor chemical potentials it is convenient to write the
result in the dimensional-reduction notation of Ref.~\cite{Vuorinen:2003fs},
\begin{equation}
 \frac{m_E^2(0)}{T^2}
 =g^2\alpha_{E4}+\frac{g^4}{(4\pi)^2}\alpha_{E6}+\mathcal{O}(g^6),
 \label{eq:VuorinenStructure}
\end{equation}
with
\begin{equation}
 \alpha_{E4}=\frac{1}{3N_f}\sum_f
 \left[\CA+\TF(1+12\bar\mu_f^2)\right]
 \label{eq:alphaE4ours}
\end{equation}
and
\begin{align}
 \alpha_{E6}&=\frac1{9N_f}\sum_f\Bigg\{
 \CA^2[22(L+\gamma_E)+5] \notag\\
 &\!\!+ \CA\TF\Big[2(7+132\bar\mu_f^2)(L+\gamma_E)+9+132\bar\mu_f^2\notag\\
 & \!\!+8\gamma_E+4\{\psi(z_f)+\psi(z_f^*)\}\Big]-18\CF\TF(1+12\bar\mu_f^2)\notag\\
 &\!\! - 4\TF^2(1+12\bar\mu_f^2)
 \Big[2L-1-\{\psi(z_f)+\psi(z_f^*)\}\Big]\Bigg\}.
 \label{eq:alphaE6ours}
\end{align}
%
Equivalently,
\begin{align}
 \frac{m_E^2(0)}{T^2}
 ={}&\frac{g^2}{3}\left[\CA+\TF\langle1+12\bar\mu_f^2\rangle_f\right]\notag\\
 &+\frac{g^4}{(4\pi)^2}\alpha_{E6}+\mathcal{O}(g^6).
 \label{eq:zeroFieldFiniteMu}
\end{align}
where $\langle X_f\rangle_f=N_f^{-1}\sum_fX_f$ and $\TF=N_f/2$.
For later use we denote the zero-field $g^4$ coefficient by
$C_0(\{\mu_f/T\};\Lambda)\equiv\alpha_{E6}/(4\pi)^2$; for a common chemical
potential we write simply $C_0(\mu/T;\Lambda)$.

\section{One-loop finite-field ingredients}
\label{sec:oneLoopB}
At nonzero magnetic field only the quark bubble acquires electromagnetic-field
dependence. We keep the full Landau spectrum and take the static limit only
after writing the exact bubble.

\subsection{Ritus propagator and the magnetized quark bubble}
We consider a homogeneous magnetic field directed along the $z$ axis and define
\begin{equation}
 b_f\equiv|q_fB|,\ \  \ s_f\equiv\operatorname{sgn}(q_fB),\  \ \ 
 \mathcal P_\pm^{(f)}=\frac{1\pm i s_f\gamma_1\gamma_2}{2}.
 \label{eq:RitusSpinProjectors}
\end{equation}
In a Ritus basis the massless level-space quark propagator is written as
\begin{equation}
 S_{f,\ell}(K_0,p_z)=\frac{-i\gamma\cdot\bar K_\ell}{D_\ell(K_0,p_z)}.
 \label{eq:RituspropMain}
\end{equation}
The denominator of the level-space propagator is
\begin{equation}
 D_\ell(K_0,p_z)=K_0^2+p_z^2+2\ell b_f.
 \label{eq:Ritusdenom}
\end{equation}
The corresponding Ritus momentum entering the numerator is
\begin{equation}
 \bar K_\ell=(K_0,0,-s_f\sqrt{2b_f\ell},p_z).
 \label{eq:RitusMomentum}
\end{equation}
For a temporal gluon carrying transverse momentum $\bm k_\perp$, the Ritus
vertex connecting Landau levels $\ell$ and $r$ is
\begin{equation}
 \mathcal V^0_{\ell r}(\bm k_\perp)
 =\gamma^0\!\left[
 F_{\ell r}(\bm k_\perp)\mathcal P_+^{(f)}
 +F_{\ell-1,r-1}(\bm k_\perp)\mathcal P_-^{(f)}\right],
 \label{eq:RitusVertexOneLoop}
\end{equation}
where an overlap with a negative Landau-level index is understood to vanish.
Only the radial modulus is needed below; with
$u\equiv k_\perp^2/(2b_f)$ it is
\begin{align}
 |F_{\ell r}(u)|^2
 ={}&e^{-u}
 \frac{\min(\ell,r)!}{\max(\ell,r)!}
 u^{|\ell-r|}
 \times
 \left[
 L_{\min(\ell,r)}^{|\ell-r|}(u)
 \right]^2.
 \label{eq:LandauOverlapMain}
\end{align}
At $k_\perp=0$, $F_{\ell r}(0)=\delta_{\ell r}$.

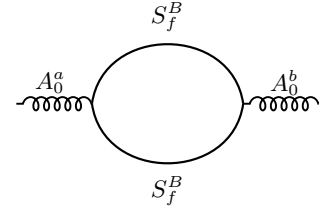
\begin{figure}[htbp]
\centering
\begin{tikzpicture}[scale=1.0]
 \draw[gluon] (-2.0,0)--(-1.0,0);
 \draw[gluon] (1.0,0)--(2.0,0);
 \draw[qline] (-1.0,0) .. controls (-.85,1.05) and (.85,1.05) .. (1.0,0);
 \draw[qline] (1.0,0) .. controls (.85,-1.05) and (-.85,-1.05) .. (-1.0,0);
 \node at (0,1.15) {$S_f^{B}$};
 \node at (0,-1.15) {$S_f^{B}$};
 \node[font=\small] at (-1.55,.28) {$A_0^a$};
 \node[font=\small] at (1.55,.28) {$A_0^b$};
\end{tikzpicture}
\caption{One-loop magnetized quark contribution to the static temporal gluon
self-energy.}
\label{fig:oneLoopB}
\end{figure}

The magnetized one-loop quark bubble is shown in
Fig.~\ref{fig:oneLoopB}.

The exact quark bubble can be written as
\begin{align}
 \Pi_{00}^{(1),q}(0,\bm k;B)
 &=-g^2\TR\sum_f\frac{b_f}{2\pi}T
 \sum_n\sum_{\ell,r\ge0}\int\frac{\dd p_z}{2\pi}\notag\\
 &\times
 \frac{\mathcal T_{\ell r}(K_n,p_z;\bm k)}
 {D_\ell(K_n,p_z)D_r(K_n,p_z+k_z)},
 \label{eq:bubbleBmain}\\
 \intertext{The numerator is the corresponding Dirac trace,}
&\hspace*{-2.15cm} \mathcal T_{\ell r}
 =\tr_D\!\left[
 \mathcal V^0_{\ell r}(\bm k_\perp)
 (-i\gamma\cdot\bar K_r')
 \mathcal V^0_{r\ell}(-\bm k_\perp)
 (-i\gamma\cdot\bar K_\ell)\right].
 \label{eq:bubbleBtraceMain}
\end{align}
Here $K_n=(2n+1)\pi T-i\mu_f$. In the second quark propagator we define
$\bar K_r'\equiv(K_n,0,-s_f\sqrt{2b_fr},p_z+k_z)$, so that its longitudinal
momentum is $p_z+k_z$ and its denominator is $D_r(K_n,p_z+k_z)$. The factor
$b_f/(2\pi)$ is the Landau degeneracy.

\subsection{Static zero-momentum reduction}
For $\bm k=0$ the temporal vertex preserves the Landau level. Oscillator
orthogonality gives $r=\ell$, and the lowest level has one spin state whereas
all $\ell\ge1$ levels have two. Defining
\begin{equation}
 \alpha_\ell\equiv2-\delta_{\ell0},\qquad
 E_{f\ell}=\sqrt{p_z^2+2\ell b_f},
\end{equation}
the Dirac trace becomes
\begin{equation}
 \mathcal T_{\ell r}(K_n,p_z;\bm0)
 =\delta_{\ell r}\,2\alpha_\ell(E_{f\ell}^2-K_n^2),
 \label{eq:Btracezero}
\end{equation}
and therefore
\begin{align}
 \Pi_{00}^{(1),q}(0,\bm0;B)
 &=-2g^2\TR\sum_f\frac{b_f}{2\pi}
 \sum_{\ell=0}^{\infty}\alpha_\ell
 \int_{-\infty}^{\infty}\frac{\dd p_z}{2\pi}\notag\\
 &\qquad\times T\sum_{n\in\mathbb Z}
 \frac{E_{f\ell}^2-K_n^2}{(K_n^2+E_{f\ell}^2)^2}.
 \label{eq:BbubbleBeforeSum}
\end{align}
The shifted fermionic Matsubara sum obeys
\begin{equation}
 T\sum_n\frac1{K_n^2+E^2}
 =\frac{1-n_F(E-\mu_f)-n_F(E+\mu_f)}{2E}.
 \label{eq:FMatsubaraBasic}
\end{equation}
We use
\begin{equation}
 -\frac{\partial}{\partial E}\left(\frac{E}{K_n^2+E^2}\right)
 =\frac{E^2-K_n^2}{(K_n^2+E^2)^2},
\end{equation}
to obtain
\begin{align}
 T\sum_n\frac{E^2-K_n^2}
 {(K_n^2+E^2)^2}
 &=-\frac1{2T}\Big[
 n_F(E-\mu_f)
  \big(1-n_F(E-\mu_f)\big)
 \notag\\
 &\hspace{-0.5cm}+n_F(E+\mu_f)
  \big(1-n_F(E+\mu_f)\big)
 \Big].
 \label{eq:FMatsubaraDerivative}
\end{align}
The vacuum term disappears under the derivative. Substitution into
Eq.~\eqref{eq:BbubbleBeforeSum} gives the exact one-loop susceptibility
 \begin{align}
 \chi_f(B)=\frac{b_f}{2\pi}&\sum_{\ell=0}^\infty\alpha_\ell
 \int_{-\infty}^{\infty}\frac{\dd p_z}{2\pi T}\notag\\
 &\times\Big[
 n_F(E_{f\ell}-\mu_f)
 (1-n_F(E_{f\ell}-\mu_f))\notag\\
 &+n_F(E_{f\ell}+\mu_f)
 (1-n_F(E_{f\ell}+\mu_f))\Big].
 \label{eq:chiBpub}
 \end{align}
The coefficient $\mathcal A_B$ defined in
Eq.~\eqref{eq:oneLoopCoefficientDefinitions} is therefore
\begin{equation}
 \mathcal A_B
 =\frac{\CA T^2}{3}+\TR\sum_f\chi_f(B).
 \label{eq:ABpub}
\end{equation}
The zero-field limit follows from
\begin{equation}
 \frac{b_f}{2\pi}\sum_{\ell\ge0}\alpha_\ell F(2b_f\ell)
 \xrightarrow[b_f\to0]{}
 2\int\frac{\dd^2p_\perp}{(2\pi)^2}F(p_\perp^2),
 \label{eq:Bto0measure}
\end{equation}
which yields
$\chi_f(0)=T^2/3+\mu_f^2/\pi^2$ and reproduces
Eq.~\eqref{eq:A0}. An equivalent Schwinger proper-time representation of
the one-loop susceptibility is derived in
Appendix~\ref{app:propagator}. There the Landau levels can be summed
exactly, producing the characteristic $\coth$ factor of the proper-time
form.

\subsection{Directional temporal kinetic coefficients}
At the next order in external momentum the magnetic field resolves the
parallel and transverse directions. The renormalized coefficients
$z_\parallel(B)$ and $z_\perp(B)$ were defined already in
Eq.~\eqref{eq:oneLoopCoefficientDefinitions}. Rotational invariance at
$B=0$ gives $z_\parallel(0)=z_\perp(0)=z_0$, with $z_0$ the zero-field
coefficient introduced in Sec.~\ref{sec:zeroB}. Only the quark sector
contributes to the magnetic differences of the directional kinetic
coefficients, which can be evaluated directly as
\begin{align}
 \Delta_B z_s
 &=\frac{1}{g^2}\frac{\partial}{\partial k_s^2}
 \Big[
 \Pi_{00}^{(1),q}(0,\bm k;B)
\left. -\Pi_{00}^{(1),q}(0,\bm k;0)
 \Big]
 \right|_{\bm k=0},
 \label{eq:DeltaZdefpub}
\end{align}
Here and throughout, $\Delta_B X\equiv X(B)-X(0)$ for a
field-dependent quantity $X$. In particular,
$\Delta_B z_s=z_s(B)-z_0$, with $s=\parallel,\perp$.

The ultraviolet pole is independent of the external magnetic field: by
ultraviolet locality it renormalizes the ordinary dimension-four gluon kinetic
operator, whereas magnetic-field-dependent corrections begin with
higher-dimensional operators. Consequently the common $\overline{\rm MS}$
pole cancels in the difference in Eq.~\eqref{eq:DeltaZdefpub}. The zero-field
coefficient was defined in Eq.~\eqref{eq:zeroBfiniteSplit} as the renormalized
quantity $z_0=W_0/(4\pi)^2$. For independent flavor chemical potentials this
is
\begin{align}
 z_0
 &=\frac{1}{(4\pi)^2}\Bigg\{
 -\frac{5\CA}{3}\left[-\frac{1}{5}+2\gamma_E
 +2\ln\frac{\Lambda}{4\pi T}\right]\notag\\
 &\hspace{0cm}+\frac{4\TR}{3}\sum_f
 \left[-1+2\ln\frac{\Lambda}{4\pi T}
 -2\Re\psi\left(\frac12-\frac{i\mu_f}{2\pi T}\right)\right]
 \Bigg\}.
 \label{eq:z0pub}
\end{align}
Here $\Re$ denotes the real part and $\psi$ is the digamma function.

An explicit finite proper-time representation for
$\Delta_B z_\parallel$ and $\Delta_B z_\perp$ is collected in
Appendix~\ref{app:directionalZ}. The zero-momentum coefficient $\mathcal A_B$
is common to both directions; anisotropy enters only through the $\mathcal{O}(k^2)$
coefficients $z_\parallel$ and $z_\perp$. Their angular combination
$z_\theta$ was defined in Eq.~\eqref{eq:ZthetaEFT}.

\subsection{Antisymmetric temporal-spatial mixing}
\label{sec:mixing}
The hard quark determinant also generates, in a commuting static color
background, the leading derivative operator
\begin{align}
 \mathcal L_{\rm mix}^{q}
 &=-ig^2\sum_f\mathcal H_f\,
 \tr_c\!\left[
 A_0(\partial_1A_2-\partial_2A_1)
 \right],
 \label{eq:localmixpub}\\
 \intertext{The flavor coefficient is given by the density response,}
 \mathcal H_f
 &=\frac{\partial n_f}{\partial(q_fB)}.
 \label{eq:mixingHfDefinition}
\end{align}
Using $\tr_c(t^at^b)=\delta^{ab}/2$ and the definition of $\mathcal H_f$
in Eq.~\eqref{eq:mixingHfDefinition}, the coefficient introduced in
Eq.~\eqref{eq:mixSelfEnergyDefinition} is
\begin{equation}
 \mathcal M(B)=\frac12\sum_f\mathcal H_f,
 \label{eq:mixingCoeffPub}
\end{equation}
which has mass dimension one. For massless flavors with a common chemical
potential, the weak-field expansion has the flavor-independent form
\begin{equation}
 \mathcal H_f
 =h_1(\mu/T)\frac{q_fB}{T}
 +h_3(\mu/T)\frac{(q_fB)^3}{T^5}+\cdots .
 \label{eq:mixingChargeExpansion}
\end{equation}
For the physical charges $q_u+q_d+q_s=0$, the leading term cancels
in the flavor sum. Higher-order magnetic terms, however, need not cancel,
particularly at finite $eB/T^2$. We therefore retain the full Landau-level
expression for $\mathcal M$ and check its size numerically over the studied
range, including $eB/T^2=4$. The off-diagonal inverse-propagator entry is
proportional to $g^2\mathcal M$.
This term is the static, long-wavelength limit of the antisymmetric part of
the fermion-loop gauge-boson polarization tensor. In momentum space it gives,
up to the Euclidean Fourier-sign convention,
\begin{equation}
 \Pi_{0i}^{ab}(0,\bm k)\big|_{\rm A}
 =-\delta^{ab}\,g^2\mathcal M\,\epsilon_{ij3}k_j+\mathcal{O}(k^3).
 \label{eq:antisymmetricPi0i}
\end{equation}
Here $i,j=1,2$, and the corresponding $i0$ component carries the opposite sign. Thus $\mathcal M$ is fixed by the coefficient of the $C$-odd antisymmetric $0i$
response in the static small-momentum limit. The thermodynamic relation
in Eq.~\eqref{eq:mixingHfDefinition} follows because the same coefficient
is the mixed derivative of the one-loop quark effective action with respect
to the temporal color source and the magnetic field.

General analyses of fermion-loop polarization in a magnetized plasma contain antisymmetric tensor structures of this type. At nonzero density they are allowed because charge conjugation is broken; in particular, the explicit finite-density magnetized polarization tensor of Wang and Shovkovy shows that
the antisymmetric form factors vanish at $\mu=0$ but are nonzero for
$\mu\ne0$~\cite{Wang:2021eud}. Their explicit calculation is for the electromagnetic polarization tensor. For the
one-loop quark diagram, however, the Dirac/Lorentz tensor structure carries over to the gluon polarization, while the electromagnetic vertex factor is replaced by the corresponding color factor; the magnetized gluon-polarization construction is discussed in Ref.~\cite{Ayala:2019akk}. Consistently, $n_f$ is odd
in $\mu_f$, so $\mathcal H_f$ and hence $\mathcal M$ vanish at $\mu_f=0$.

For $\bm k=k(\sin\theta,0,\cos\theta)$,
Eq.~\eqref{eq:antisymmetricPi0i} mixes $A_0$ with $A_y$ with amplitude
proportional to $g^2\mathcal M k\sin\theta$. Since $\mathcal M$ is a hard one-loop coefficient of mass dimension one,
its Schur-complement contribution is $\mathcal{O}(g^4T^2)$.

\section{Two-loop self-energy at arbitrary magnetic field}
\label{sec:twoLoopB}
We now turn to the $B$-dependent $g^4$ term. Pure-glue and ghost two-loop
graphs are independent of the external electromagnetic field and therefore
drop out of the magnetic difference. The five quark-bearing topologies in
Fig.~\ref{fig:twoLoopZero} are generated together by differentiating the
quark-gluon exchange functional twice with respect to the temporal source.
The two QED-like graphs form the $qq$ class, the mixed non-Abelian graph the
$qg$ class, and the two polarization-insertion graphs the $gg$ class. The
magnetic field changes the representation of this exchange functional, not
its diagrammatic organization.

\subsection{Exchange generator}
We use the temporal color source and shifted-frequency variable
introduced in Eq.~\eqref{eq:zeroBbackground}, without redefining them.
As at zero field, both quark lines and the exchanged gluon carry the
corresponding source shifts; at finite $B$ the quark propagators and vertices
are represented by their exact magnetized Ritus forms. The shifted energies
are
\begin{align}
 K_{ni}&=\omega_n-i\mu_f+\varphi h_i,
 \notag\\
 Q_{mj}&=\omega_m-i\mu_f+\varphi h_j,
 \notag\\
 R_{nm,ij}&=K_{ni}-Q_{mj}.
 \label{eq:BsourceShifts}
\end{align}
where $\omega_n=(2n+1)\pi T$.

The finite-field calculation uses the same two-loop vacuum exchange
skeleton as Eq.~\eqref{eq:zeroBexchangeFunctional}, shown in
Fig.~\ref{fig:exchangeVacuumSkeleton}, with the quark propagators replaced
by their magnetized counterparts. We denote its contribution by
$\mathscr G_{\rm exch}(\varphi,B)$, using the same sign convention as at
zero field: $\Gamma_{\rm exch}/(\beta V)=-\mathscr G_{\rm exch}$.
The second derivative with respect to $\varphi$ generates the same
quark-bearing self-energy topologies as in the zero-field calculation.

More explicitly, we obtain this exchange graph by expanding the Euclidean interaction to second order and
contracting the two gluon fields. At fixed
temporal source, the corresponding coordinate-space expression for
$\mathscr G_{\rm exch}$ is
\begin{align}
 \mathscr G_{\rm exch}(\varphi,B)
 ={}&\frac{g^2}{2}\sum_f\int_{x,y}
 D_{\mu\nu}^{ab}(x-y;\varphi)
 \notag\\
 &\hspace{-2.7cm}\times
 \tr_{D,c}\!\left[
 \gamma_\mu t^a S_f^B(x,y;\varphi)
 \gamma_\nu t^b S_f^B(y,x;\varphi)
 \right].
 \label{eq:exchangeCoordinateFull}
\end{align}
where $D_{\mu\nu}^{ab}(x-y;\varphi)$ is the internal gluon propagator in the same
constant temporal source. The temporal source acts on both charged quark
propagators and on the color-charged internal gluon propagator. Consequently its second derivative
generates both the QED-like and non-Abelian self-energy insertions.

In a homogeneous electromagnetic background the coordinate-space quark
propagator factorizes into a Schwinger phase and a translationally invariant
core,
\begin{equation}
 S_f^B(x,y;\varphi)=e^{i\Phi_f(x,y)}\,\widetilde S_f^B(x-y;\varphi),
 \label{eq:SchwingerPhaseFactorization}
\end{equation}
The Schwinger phase is antisymmetric under interchange of its endpoints,
\begin{equation}
 \Phi_f(y,x)=-\Phi_f(x,y),
 \label{eq:SchwingerPhaseAntisymmetry}
\end{equation}
Here $\widetilde S_f^B$ denotes the translationally invariant core containing
the exact Landau-level structure.
The exchange skeleton contains the product
\begin{equation}
 S_f^B(x,y;\varphi)S_f^B(y,x;\varphi)
 =\widetilde S_f^B(x-y;\varphi)\widetilde S_f^B(y-x;\varphi),
 \label{eq:SchwingerPhaseCancellation}
\end{equation}
because
$e^{i\Phi_f(x,y)}e^{i\Phi_f(y,x)}=1$. Thus the electromagnetic phase cancels exactly around the closed quark loop,
while the internal gluon carries no electromagnetic phase. We therefore work
with the translationally invariant propagator cores after
Eq.~\eqref{eq:SchwingerPhaseCancellation}; the full $B$ dependence remains in
$b_f=|q_fB|$, the Landau energies, spin projectors, and the Ritus/Landau
overlap functions.

Here $\int_{x,y}$ in Eq.~\eqref{eq:exchangeCoordinateFull} denotes the two
thermal spacetime integrations. Using
Eq.~\eqref{eq:SchwingerPhaseCancellation}, the Schwinger phases in the two
quark propagators cancel before any Fourier/Ritus transformation is made.
The exchange contribution therefore becomes an ordinary momentum/Ritus-space
integral. For the source-shifted internal gluon we use background-field Feynman gauge
for the quantum field and differentiate only at $\varphi=0$. The resulting
background-field curvature is converted analytically to the ordinary
Feynman-gauge two-point function before the screening pole is assembled. The
translationally invariant expression is
\begin{widetext}
\begin{align}
 \mathscr G_{\rm exch}(\varphi,B)
 &=\frac{g^2}{2}\sum_f\frac{b_f}{2\pi}
 \sum_{i,j=1}^{N_c}\mathcal C_{ij}\,T^2
 \sum_{n,m}\sum_{\ell,r\ge0}\notag\\
 &\quad\times
 \int\frac{\dd p_z\dd q_z}{(2\pi)^2}
 \int\frac{\dd^2k_\perp}{(2\pi)^2}
 \frac{\mathcal N_{\ell r}(K_{ni},Q_{mj};\bm k_\perp)}
 {D_\ell(K_{ni},p_z)D_r(Q_{mj},q_z)
  G(R_{nm,ij},p_z-q_z,k_\perp)}.
 \label{eq:exchangefunctionalpub}\\
 \intertext{Here the color weight associated with the exchanged gluon is}
 \mathcal C_{ij}&=\sum_a(t^a)_{ij}(t^a)_{ji}
 =\TR\left(1-\frac{\delta_{ij}}{N_c}\right),\\
 \intertext{while the source-shifted gluon denominator, including the auxiliary infrared regulator, is}
 G(R,p_z-q_z,k_\perp)&=R^2+(p_z-q_z)^2+k_\perp^2+\lambda_{\rm IR}^2.
 \label{eq:colorweightpub}
\end{align}
\end{widetext}
The auxiliary infrared regulator $\lambda_{\rm IR}$ is used only to organize the static exchanged-
gluon region; it is removed by the strict-DR prescription below. Equation~\eqref{eq:exchangefunctionalpub} is therefore written after the
cancelling overall Schwinger phases have been removed, while the complete
Landau-level and transverse-overlap dependence is retained.

\subsection{Exact Landau-level numerator}
For $u=k_\perp^2/(2b_f)$, we define the real radial overlap
\begin{align}
 f_{ab}(u)
 ={}&e^{-u/2}
 \sqrt{\frac{\min(a,b)!}{\max(a,b)!}}
 \notag\\
 &\times
 u^{|a-b|/2}
 L_{\min(a,b)}^{|a-b|}(u).
 \label{eq:radialOverlapMain}\\
 \intertext{For negative Landau-level indices, the overlap is set to zero:}
 f_{ab}&=0
 \qquad (a<0\ \text{or}\ b<0).
 \label{eq:radialOverlapNegativeIndices}
\end{align}
The temporal Ritus vertex $\mathcal V^0_{\ell r}$ was defined in
Eq.~\eqref{eq:RitusVertexOneLoop}. Using the same overlap functions and
the spin projectors of Eq.~\eqref{eq:RitusSpinProjectors}, the remaining
spatial vertices needed for the two-loop exchange are
\begin{align}
 \mathcal V_{\ell r}^{3}
 &=\gamma^{3}[F_{\ell r}\mathcal P_+^{(f)}
 +F_{\ell-1,r-1}\mathcal P_-^{(f)}],\notag\\
 \mathcal V_{\ell r}^{1,2}
 &=\gamma^{1,2}[F_{\ell-1,r}\mathcal P_+^{(f)}
 +F_{\ell,r-1}\mathcal P_-^{(f)}].
 \label{eq:RitusVerticesTwoLoop}
\end{align}
The massless ordered Dirac trace summed over the exchanged-gluon Lorentz index
reduces to
\begin{equation}
 \mathcal N_{\ell r}
 =A_{\ell r}(K_0Q_0+p_zq_z)+C_{\ell r},
 \label{eq:twoLoopTraceMain}
\end{equation}
with
\begin{align}
 A_{\ell r}&=4[f_{\ell,r-1}(u)^2+f_{\ell-1,r}(u)^2],\notag\\
 C_{\ell r}&=16b_f\sqrt{\ell r}\,
 f_{\ell r}(u)f_{\ell-1,r-1}(u).
 \label{eq:twoLoopACMain}
\end{align}
This is an exact full-level expression; no expansion in $b_f/T^2$ or in
$u$ has been made. In particular, the full vertex is not diagonal in the
Landau level inside the two-loop exchange graph.

\subsection{Temporal-source representation of the two-loop self-energy}
Exactly as in the zero-field construction of
Eq.~\eqref{eq:zeroBPi2FromI0}, the static two-loop exchange curvature is
obtained by differentiating the source-shifted exchange functional twice
with respect to $\varphi$, followed by setting $\varphi=0$. The color
projection along the normalized generator $\mathsf H$ is understood.
Each $\varphi$ derivative inserts a zero-momentum temporal source;
$\partial_a^2=g^2\partial_\varphi^2$ supplies the vertex factors
associated with the physical field amplitude. The difference from the zero-field case
lies in the magnetized quark propagators and vertices, not in the
source-curvature prescription.

The frequency-shift differentiation is a background-field construction. In
background-field Feynman gauge the color-charged exchanged gluon carries the
source shift $R_0\to R_0+\varphi(h_i-h_j)$, so differentiating its propagator
produces the background three- and four-gluon source vertices. This is not
identical, diagram by diagram, to differentiating an ordinary covariant
Feynman-gauge propagator. In particular, for one zero-momentum temporal leg
the ordinary color-stripped three-gluon vertex is
\begin{equation}
 V^{\rm F}_{0\alpha\beta}(0,R,-R)
 =2R_0\delta_{\alpha\beta}
 -\delta_{0\alpha}R_\beta-\delta_{0\beta}R_\alpha,
 \label{eq:zeroMomentumThreeGluon}
\end{equation}
whereas the background-field Feynman-gauge source derivative generates only
the $2R_0\delta_{\alpha\beta}$ part. At first source order the longitudinal
pieces vanish after contraction with the uninserted conserved quark current.
At second order this argument no longer applies to the mixed $qg$ sector,
because a quark current carrying one temporal source insertion is not itself
conserved. The longitudinal three-gluon pieces must therefore be retained
until the $qg$ and $gg$ sectors are combined. Their $gg$ remainder cancels
against part of the $qg$ remainder, leaving the finite local conversion
derived below. Thus the shifted-frequency kernels used in the numerical
calculation are background-field Feynman-gauge curvatures, and the ordinary
Feynman-gauge self-energy is obtained only after this analytic conversion.

For the magnetized exchange functional in
Eq.~\eqref{eq:exchangefunctionalpub}, the corresponding unrenormalized
background-field Feynman-gauge curvature is therefore
\begin{equation}
 \Pi_{00,\rm BF,bare}^{(2),q}(0,\bm0;B)
 =-g^2\left.\frac{\partial^2}{\partial\varphi^2}
 \mathscr G_{\rm exch}(\varphi,B)\right|_{\varphi=0},
 \label{eq:Pi2sourceMain}
\end{equation}
which is the finite-$B$ counterpart of Eq.~\eqref{eq:zeroBPi2FromI0}.
We do not introduce a separate counterterm functional for this auxiliary
curvature. Instead, its magnetic difference is first ultraviolet-subtracted
with the common zero-field forest, then converted analytically to the ordinary
Feynman-gauge two-point function through
Eq.~\eqref{eq:zeroBbgToFeynman}. Only the converted quantity is
inserted into the strict pole relation
Eq.~\eqref{eq:screeningFromSelfEnergy}, where the bare QCD coupling is
expressed through Eq.~\eqref{eq:zeroBgct}. The external electromagnetic field
does not create a new QCD ultraviolet counterterm.

The diagrammatic meaning of the differentiation can be seen directly from
matrix-inverse identities. If $S=\mathcal D_q^{-1}$ is a quark propagator in
the presence of the temporal source and $D=\mathcal D_g^{-1}$ is the internal gluon propagator,
then
\begin{subequations}\label{eq:propagatorSourceDerivativeIdentity}
\begin{align}
 \partial_\varphi S&=-S(\partial_\varphi\mathcal D_q)S,
 \label{eq:propagatorSourceDerivativeS}\\
 \partial_\varphi D&=-D(\partial_\varphi\mathcal D_g)D,
 \label{eq:propagatorSourceDerivativeD}\\
 \partial_\varphi^2D&={}2D(\partial_\varphi\mathcal D_g)D
 (\partial_\varphi\mathcal D_g)D-D(\partial_\varphi^2\mathcal D_g)D.
 \label{eq:propagatorSourceDerivativeD2}
\end{align}
\end{subequations}
Thus a derivative of a quark propagator inserts one zero-momentum temporal
quark-gluon vertex on that line. For the internal gluon, the first derivative
inserts the source-induced three-gluon vertex. The two terms in
$\partial_\varphi^2D$ have the familiar non-Abelian interpretation: the first gives
two successive three-gluon insertions, while the second gives the four-gluon
contact insertion. Applying the product rule to the two quark propagators and
the one internal gluon in Eq.~\eqref{eq:exchangeCoordinateFull} therefore
produces all inequivalent positions of two external $A_0$ legs. The result is
precisely a two-point 1PI vertex at zero external four-momentum, i.e. the
static temporal self-energy. The correspondence with the individual graphs
follows directly from the placement of the two source derivatives.

The differentiation operator is the same as in the zero-field case:
\begin{equation}
 \frac{\partial}{\partial\varphi}
 =h_i\frac{\partial}{\partial K_0}
 +h_j\frac{\partial}{\partial Q_0}
 +(h_i-h_j)\frac{\partial}{\partial R_0}.
 \label{eq:sourceDerivativeOperator}
\end{equation}
Here $K_0$, $Q_0$, and $R_0$ are independent kernel arguments during
differentiation, with $R_0=K_0-Q_0$ imposed afterwards. The exchange
skeleton carries $g^2$, and the overall $g^2$ in
Eq.~\eqref{eq:Pi2sourceMain} supplies the two additional powers of the
coupling. Equations~\eqref{eq:propagatorSourceDerivativeIdentity} and~\eqref{eq:sourceDerivativeOperator} generate the five two-loop topologies
with the same combinatorial organization as at $B=0$.

To display the result algebraically, set
\begin{align}
 \mathscr H_{\ell r}(K_0,Q_0)
 &\equiv
 \frac{\mathcal N_{\ell r}}
 {D_\ell(K_0)D_r(Q_0)},
 \notag\\
 \mathcal Q_{ij}
 &\equiv
 h_i\partial_{K_0}
 +h_j\partial_{Q_0},
 \notag\\
 \Delta h_{ij}
 &\equiv h_i-h_j.
 \label{eq:HoperatorMain}
\end{align}
Applying the same source-derivative identity as at zero field separates the
curvature into $qq$, $qg$, and $gg$ sectors according to whether the two
insertions act on the quark lines, one quark line and the exchanged gluon, or
the exchanged gluon twice. For numerical implementation we carry out the
quark-line derivatives explicitly and define
\begin{subequations}\label{eq:twoLoopAbbrevMain}
\begin{align}
 U_{\ell r}&=\frac1{D_\ell D_r},
 \label{eq:UlrMain}\\
 x_q&=\frac{h_iK_0}{D_\ell}+\frac{h_jQ_0}{D_r},
 \label{eq:xqMain}\\
 y_q&={}h_i^2\left(\frac1{D_\ell}-\frac{2K_0^2}{D_\ell^2}\right)\notag\\
 &\quad+h_j^2\left(\frac1{D_r}-\frac{2Q_0^2}{D_r^2}\right),
 \label{eq:yqMain}\\
 d_N&=A_{\ell r}(h_iQ_0+h_jK_0),\qquad
 N\equiv\mathcal N_{\ell r}.
 \label{eq:dNMain}
\end{align}
\end{subequations}
Then the pointwise kernels are
\begin{subequations}\label{eq:pointwiseKernelsMain}
\begin{align}
 \mathcal K_{qq}&=\frac{U_{\ell r}}{G}
 \left[2h_ih_jA_{\ell r}-4x_qd_N+(4x_q^2-2y_q)N\right],
 \label{eq:KqqExplicitMain}\\
 \mathcal K_{qg}&=-\frac{4\Delta h_{ij}R_0U_{\ell r}}{G^2}
 (d_N-2x_qN),
 \label{eq:KqgExplicitMain}\\
 \mathcal K_{gg}&=(\Delta h_{ij})^2U_{\ell r}N
 \left(\frac{8R_0^2}{G^3}-\frac2{G^2}\right).
 \label{eq:KggExplicitMain}
\end{align}
\end{subequations}
The bare finite-field exchange curvature is therefore
\begin{align}
 \Pi_{00,\rm BF,bare}^{(2),q}(0,\bm0;B)
 ={}&-\frac{g^4T^2}{2}
 \sum_f\frac{b_f}{2\pi}
 \sum_{i,j}\mathcal C_{ij}\notag\\
 &\hspace{-3.8em}\times\sum_{n,m}\sum_{\ell,r\ge0}\int_{p,q,k}
 [\mathcal K_{qq}+\mathcal K_{qg}+\mathcal K_{gg}],
 \label{eq:Pi2Bexplicit}
\end{align}
where
$\int_{p,q,k}\equiv\int\dd p_z\dd q_z/(2\pi)^2
\int\dd^2k_\perp/(2\pi)^2$.
The three sectors are not separately renormalizable Feynman graphs; only their
sum is the source derivative of the gauge-complete exchange kernel. In the
$gg$ sector, the $8R_0^2/G^3$ and $-2/G^2$ terms arise from the sequential
three-gluon and quadratic four-gluon source insertions, respectively.
Equation~\eqref{eq:Pi2Bexplicit} gives the bare finite-field
background-field Feynman-gauge exchange curvature. Its conversion to the
ordinary covariant-Feynman self-energy follows the same local relation as at
zero field, Eq.~\eqref{eq:zeroBbgToFeynman}, with the one-loop quark
self-energy now evaluated in the magnetic background. The conversion is applied to the magnetic difference after the ultraviolet
forest subtraction. Its pole cancels in the assembled strict-pole expansion,
while the finite part gives the local contribution displayed below.

The generator and SU(3) color projection are those already defined
in Eqs.~\eqref{eq:sourceGeneratorH} and~\eqref{eq:colorprojectionpub}.
At vanishing source, the projected curvature is independent of the chosen
normalized color direction.

\subsection{Finite-field subtraction and analytic local pieces}
\label{sec:finiteBsubtraction}
The finite-field calculation is organized in direct analogy with the
zero-field exchange calculation. The difference is practical rather than
conceptual: at $B=0$ the exchange curvature can be reduced analytically to
standard sum-integrals, whereas at finite $B$ it is evaluated as a
multidimensional Landau/Ritus integral. For numerical stability it is useful
to subtract its local large-momentum asymptotes at the integrand level and add
them back in a lower-dimensional analytic representation.

\begin{figure*}[tbh]
\centering
\begin{tikzpicture}[scale=1.0]
 \begin{scope}[xshift=-3.2cm]
   \coordinate (L) at (-.9,0); \coordinate (Rr) at (.9,0);
   \draw[qline] (L) .. controls (-.45,.85) and (.45,.85) .. (Rr);
   \draw[qline] (Rr) .. controls (.45,-.85) and (-.45,-.85) .. (L);
   \draw[gluon] (L) .. controls (0,.18) .. (Rr);
   \draw[-{Stealth[length=1.8mm,width=1.4mm]},line width=.7pt]
        (-.28,.72)--(.05,.78);
   \draw[-{Stealth[length=1.8mm,width=1.4mm]},line width=.7pt]
        (.28,-.72)--(-.05,-.78);
   \node[font=\small] at (0,.98) {$P$};
   \node[font=\small] at (0,-.98) {$Q$ spectator};
   \node[font=\small,fill=white,inner sep=1pt] at (0,.37) {$R$};
   \draw[dashed,rounded corners=3pt] (-1.18,-.12) rectangle (1.18,1.15);
   \node[font=\footnotesize] at (0,1.38) {$P$-side: $P,R$ hard};
 \end{scope}
 \begin{scope}[xshift=3.2cm]
   \coordinate (L) at (-.9,0); \coordinate (Rr) at (.9,0);
   \draw[qline] (L) .. controls (-.45,.85) and (.45,.85) .. (Rr);
   \draw[qline] (Rr) .. controls (.45,-.85) and (-.45,-.85) .. (L);
   \draw[gluon] (L) .. controls (0,-.18) .. (Rr);
   \draw[-{Stealth[length=1.8mm,width=1.4mm]},line width=.7pt]
        (-.28,.72)--(.05,.78);
   \draw[-{Stealth[length=1.8mm,width=1.4mm]},line width=.7pt]
        (.28,-.72)--(-.05,-.78);
   \node[font=\small] at (0,.98) {$P$ spectator};
   \node[font=\small] at (0,-.98) {$Q$};
   \node[font=\small,fill=white,inner sep=1pt] at (0,-.35) {$R$};
   \draw[dashed,rounded corners=3pt] (-1.18,-1.15) rectangle (1.18,.12);
   \node[font=\footnotesize] at (0,-1.38) {$Q$-side: $Q,R$ hard};
 \end{scope}
\end{tikzpicture}
\caption{One-sided ultraviolet regions of the exchange skeleton associated
with the $P$- and $Q$-side zero-field subgraphs.}
\label{fig:UVsubgraphs}
\end{figure*}
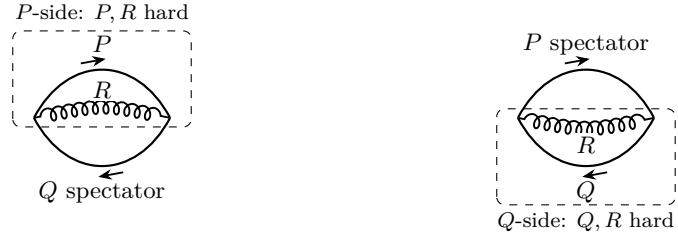

A particularly transparent form is obtained by writing the translationally
invariant magnetized quark propagator as
\begin{equation}
 \widetilde S_f^B(P;\varphi)=S_f^0(P;\varphi)+\delta S_f(P;\varphi,B).
 \label{eq:magnetizedPropagatorDifference}
\end{equation}
Here the magnetic correction vanishes continuously in the zero-field limit,
\begin{equation}
 \delta S_f(P;\varphi,B)\xrightarrow[B\to0]{}0.
 \label{eq:deltaSdefinition}
\end{equation}
For the ultraviolet subtraction we use the equivalent Schwinger-momentum
representation, so the finite- and zero-field terms are defined on one common
momentum chart. Using Eq.~\eqref{eq:magnetizedPropagatorDifference}, the
magnetic difference of the exchange functional has the exact decomposition
\begin{widetext}
\begin{align}
 \Delta_B\mathscr G_{\rm exch}(\varphi)
 =\frac{g^2}{2}\sum_f\int_{P,Q}D_{\mu\nu}(P-Q;\varphi)
 \tr_{D,c}\Big[\gamma_\mu\delta S_f(P)\gamma_\nu S_f^0(Q)
 +\gamma_\mu S_f^0(P)\gamma_\nu\delta S_f(Q)+\gamma_\mu\delta S_f(P)\gamma_\nu\delta S_f(Q)\Big] .
 \label{eq:exchangeDeltaSdecomp}
\end{align}
\end{widetext}
The first two terms contain the two one-sided zero-field quark self-energy subgraphs; the last term contains a magnetic difference on both quark lines and is ultraviolet convergent. The two one-sided regions are illustrated in
Fig.~\ref{fig:UVsubgraphs}.

The zero-field subgraph appearing in the first term of
Eq.~\eqref{eq:exchangeDeltaSdecomp} can be integrated before the remaining
magnetized line. We define
\begin{align}
 \mathcal U_0(P)
 &\equiv T\sum_{q_0}\int_{\bm q}
 \frac{\gamma_\mu S_0(Q)\gamma_\mu}
 {(P-Q)^2+\lambda_{\rm IR}^2}\notag\\
 &=\mathcal U_{0,\rm vac}^{\overline{\rm MS}}(P)
 +\mathcal U_{0,\rm mat}(P).
 \label{eq:U0definition}
\end{align}
where the auxiliary infrared regulator $\lambda_{\rm IR}$ regulates only the static exchanged
gluon and is removed in the strict-DR limit. For massless quarks the
renormalized vacuum part is proportional to $i\slashed P$,
\begin{equation}
 \mathcal U_{0,\rm vac}^{\overline{\rm MS}}(P)
 =i\slashed P\,c_{\rm vac}(P^2;\lambda_{\rm IR},\Lambda).
 \label{eq:U0vacMS}
\end{equation}
The scalar coefficient in Eq.~\eqref{eq:U0vacMS} is
\begin{align}
 c_{\rm vac}(P^2;\lambda_{\rm IR},\Lambda)
 &=\frac{2}{(4\pi)^2}\int_0^1\!\dd x\,(1-x)\notag\\
 &\hspace{.0cm} \times
 \ln\!\left[
 \frac{\Lambda^2}
 {x(1-x)P^2+(1-x)\lambda_{\rm IR}^2}
 \right].
 \label{eq:cvacExplicit}
\end{align}
Setting the auxiliary gluon mass to zero, $\lambda_{\rm IR}\to0$, gives
\begin{equation}
 c_{\rm vac}(P^2;0,\Lambda)
 =\frac{1}{(4\pi)^2}
 \left[\ln\frac{\Lambda^2}{P^2}+2\right].
 \label{eq:cvacMassless}
\end{equation}
Thus the ultraviolet-local vacuum addback is explicit.

The matter part can likewise be written after the Matsubara sum.
At vanishing source, let $E_q=|\bm q|$,
$\omega_{pq}=\sqrt{|\bm p-\bm q|^2+\lambda_{\rm IR}^2}$ and $\zeta_P=iP_0$. Then
\begin{align}
 \mathcal U_{0,\rm mat}(P)
 &=-\int_{\bm q}\sum_{s_1,s_2=\pm1}\notag\\
 &\quad\times \mathcal W_{s_1s_2}(P,Q)
 \left(\gamma_0+i\,s_1\,\frac{\bm\gamma\!\cdot\!\bm q}{E_q}\right),
 \label{eq:U0matter}\\
 \intertext{The corresponding thermal weight is}
 \mathcal W_{s_1s_2}(P,Q)
 &=\frac{s_2}{2\omega_{pq}}\notag\\
 &\quad\times
 \frac{s_2\,n_B(\omega_{pq})-s_1\,n_F(E_q-s_1\mu)}
 {\zeta_P-s_1E_q-s_2\omega_{pq}} .
 \label{eq:WstMatter}
\end{align}
Here $s_1,s_2=\pm1$ label the fermionic and bosonic energy-pole signs,
respectively.
For the source curvature, the frequency shifts of
Eq.~\eqref{eq:BsourceShifts} are inserted before differentiating. Equations~\eqref{eq:U0vacMS}-\eqref{eq:WstMatter} are the analytic one-sided addbacks
used in the numerical implementation; only the remaining momentum integrals
are performed numerically.

The vacuum piece simplifies further. Writing the exact Schwinger propagator
coefficients as
\begin{equation}
 \delta S_f(P)
 =-i\left[\delta A_f\slashed P_\parallel
 +\delta C_f\slashed P_\perp
 -H_f\slashed P_\parallel i\gamma^1\gamma^2\right],
 \label{eq:deltaSSchwinger}
\end{equation}
the proper-time integration-by-parts identity gives
\begin{equation}
 \delta A_f P_\parallel^2+\delta C_f P_\perp^2=0.
 \label{eq:SchwingerVacIdentity}
\end{equation}
Here the dimensional continuation is fixed by keeping the magnetic $(1,2)$
plane exactly two dimensional and taking
$D=4-2\epsilon$, so that the complementary parallel subspace has dimension
$D-2=2-2\epsilon$; equivalently, the spatial momentum parallel to $\bm B$
is continued to $1-2\epsilon$ dimensions. The identity in
Eq.~\eqref{eq:SchwingerVacIdentity} is valid in this continuation before
setting $\epsilon$ to zero. Indeed, with
\begin{subequations}
\begin{align}
 \mathcal E_B(\tau)&\equiv
 \exp\!\left[-\tau P_\parallel^2
 -\frac{P_\perp^2\tanh(b_f\tau)}{b_f}\right],\\
 \mathcal E_0(\tau)&\equiv
 \exp\!\left[-\tau(P_\parallel^2+P_\perp^2)\right],
\end{align}
\end{subequations}
one has
\begin{align}
 \frac{\dd}{\dd\tau}(\mathcal E_B-\mathcal E_0)
 ={}&-P_\parallel^2(\mathcal E_B-\mathcal E_0)\notag\\
 &-P_\perp^2\left[
 \mathrm{sech}^2(b_f\tau)\mathcal E_B-\mathcal E_0\right].
 \label{eq:SchwingerVacIdentityD}
\end{align}
The boundary term from integrating Eq.~\eqref{eq:SchwingerVacIdentityD}
over $\tau$ vanishes, which gives Eq.~\eqref{eq:SchwingerVacIdentity} for
arbitrary $\epsilon$. The same argument holds before the temporal source is
set to zero, with $P_0\to P_{0,i}(\varphi)$.

In the same $D$-dimensional continuation the renormalized zero-field vacuum
subgraph has the Lorentz-covariant form
$\mathcal U_{0,\rm vac}^{\overline{\rm MS}}
=i\slashed P_i(\varphi)c_{\rm vac}(P_i^2;\epsilon)$. Therefore
\begin{equation}
 \tr_D\!\left[
 \delta S_f(P_i;\varphi,B)
 \mathcal U_{0,\rm vac}^{\overline{\rm MS}}(P_i;\varphi)
 \right]=0,
 \label{eq:sourcewiseVacuumZero}
\end{equation}
for every $\varphi$ and every $\epsilon$. The terms proportional to
$\delta A_f$ and $\delta C_f$ cancel by
Eq.~\eqref{eq:SchwingerVacIdentity}, while the spin term vanishes because
$\tr_D(\slashed P_\parallel\gamma^1\gamma^2\slashed P)=0$ for the fixed
magnetic two-plane. Hence the complete source-differentiated local vacuum
addback vanishes identically before the $\epsilon\to0$ limit. In particular,
it contains no $\mathcal{O}(\epsilon)$ term that could combine with a $1/\epsilon$ pole
to generate an additional finite magnetic constant.

Taking two source derivatives of Eq.~\eqref{eq:sourcewiseVacuumZero} generates the local $qq$, $qg$, and $gg$ source-insertion pieces. Operationally, the numerical remainder is obtained by replacing each one-sided bare zero-field subgraph by its matter part $\mathcal U_{0,\rm mat}$ before taking the source curvature.

Equation~\eqref{eq:sourcewiseVacuumZero} has a stronger consequence for the
magnetic difference. The renormalized vacuum parts of the one-sided local
subgraphs vanish for every source value and every $\epsilon$, while the matter
addbacks in Eq.~\eqref{eq:U0matter} and the double-difference term in
Eq.~\eqref{eq:exchangeDeltaSdecomp} are ultraviolet finite. Therefore the
background-field Feynman-gauge magnetic difference has a regular
$\epsilon\to0$ limit. The ultraviolet pole of the ordinary Feynman-gauge
magnetic self-energy arises entirely from the local conversion
Eq.~\eqref{eq:zeroBbgToFeynman}.

To make this explicit, continue the one-loop magnetic difference in the same
dimensional scheme as
\begin{equation}
 \frac{\Delta_B\Pi_{00}^{(1),q}}{g^2T^2}
 =\Delta_BM_0+\epsilon\,\Delta_BM_1+\mathcal{O}(\epsilon^2),
 \label{eq:finiteBOneLoopDimensionalExpansion}
\end{equation}
where $\Delta_BM_0\equiv(\mathcal A_B-\mathcal A_0)/T^2
=\Delta_B\mathcal A/T^2$, with $\mathcal A_B$ and $\mathcal A_0$
defined in Eqs.~\eqref{eq:ABpub} and~\eqref{eq:A0}, respectively.
Using Eq.~\eqref{eq:zeroBsumintEval}, the magnetic part of the
background-to-ordinary conversion is then
\begin{align}
 \frac{\Delta_B\Pi_{00}^{(2),q}
       -\Delta_B\Pi_{00,\mathrm{BF}}^{(2),q}}
      {g^4T^2}
 ={}&\frac{2\CA}{(4\pi)^2}
 \Bigg[
 \frac{\Delta_BM_0}{\epsilon}
 +\Delta_BM_1\notag\\
 &\hspace{-10mm}
 +2(L+\gamma_E)\Delta_BM_0
 \Bigg]+\mathcal{O}(\epsilon).
 \label{eq:finiteBFeynmanConversionExpanded}
\end{align}
For a coefficient $X$ of mass dimension two, a hat denotes the
normalization $\widehat X\equiv X/T^2$; consequently,
$\Delta_B\widehat X=\Delta_BX/T^2$.
The origin of the strict-pole subtraction follows directly from
Eq.~\eqref{eq:zeroBassembly}. Replacing $g_0^2$ with the renormalized
coupling according to Eq.~\eqref{eq:zeroBgct} produces, in the magnetic
difference, the $\mathcal{O}(g^4T^2)$ term
$-g^4T^2\beta_0(\Delta_BM_0+\epsilon\Delta_BM_1)/
[(4\pi)^2\epsilon]$. Separately, the ultraviolet pole
$g^2W_{-1}/[(4\pi)^2\epsilon]$ of the one-loop momentum derivative
in Eq.~\eqref{eq:zeroBpoleIngredientsB} contributes
$-g^4T^2W_{-1}(\Delta_BM_0+\epsilon\Delta_BM_1)/
[(4\pi)^2\epsilon]$ through
$-\Pi_{00}^{(1)}\partial_{k^2}\Pi_{00}^{(1)}$.
Dividing the sum of these two contributions by $g^4T^2$ defines
$\Delta_B\widehat c_{\rm strict}^{\rm ct}$. Keeping the
$\mathcal{O}(\epsilon)$ part of the one-loop magnetic difference in
Eq.~\eqref{eq:finiteBOneLoopDimensionalExpansion}, which multiplies
the ultraviolet pole, gives
\begin{align}
 \Delta_B\widehat c_{\rm strict}^{\rm ct}
 &=-\frac{\beta_0+W_{-1}}{(4\pi)^2}
 \left(
 \frac{\Delta_BM_0}{\epsilon}+\Delta_BM_1
 \right)\notag\\
 &=-\frac{2\CA}{(4\pi)^2}
 \left(
 \frac{\Delta_BM_0}{\epsilon}+\Delta_BM_1
 \right).
 \label{eq:finiteBStrictCounterterm}
\end{align}
Since $\beta_0+W_{-1}=2\CA$ in ordinary Feynman gauge, the second
line follows directly. This strict-pole subtraction is not an additional
magnetic counterterm in the QCD Lagrangian. Adding
Eqs.~\eqref{eq:finiteBFeynmanConversionExpanded} and~\eqref{eq:finiteBStrictCounterterm} leaves the finite local conversion
\begin{equation}
 \frac{2\CA}{(4\pi)^2}\,
 2(L+\gamma_E)\,\Delta_BM_0.
 \label{eq:finiteBConversionFiniteRemainder}
\end{equation}
The magnetic exchange difference in Eq.~\eqref{eq:exchangeDeltaSdecomp}
is treated by the same two source derivatives used in
Eq.~\eqref{eq:Pi2sourceMain}. As in Eq.~\eqref{eq:Pi2Bexplicit}, the
resulting curvature is that of the auxiliary background-field exchange
functional. After subtracting the one-sided ultraviolet subgraphs, its
finite dimensionless magnetic remainder is
\begin{align}
 \Delta_B\widehat c_{\rm rem}^{(3)}
 ={}&-\frac{1}{g^2T^2}
 \lim_{\lambda_{\rm IR}\to0}^{\rm strict\,DR}\notag\\
 &\hspace{-9mm}\times\left[\mathcal R_{\rm loc}
 \left(\left.\frac{\partial^2}{\partial\varphi^2}
 \Delta_B\mathscr G_{\rm exch}(\varphi;\lambda_{\rm IR})
 \right|_{\varphi=0}\right)\right]_{d=3}.
 \label{eq:finiteBremFromFunctional}
\end{align}
Here $\mathcal R_{\rm loc}$ removes the source-differentiated
ultraviolet forest before the sum-first finite
remainder is evaluated at $d=3$ spatial dimensions. The superscript
$(3)$ denotes the spatial dimension, not the order in $g$; the static
exchanged-gluon overlap is treated by the strict-DR prescription below.
Equation~\eqref{eq:numericalCremExplicit} gives the corresponding explicit
Landau/Ritus integral. The apparently evanescent zero-field thermal term
in Eqs.~\eqref{eq:zeroBbgPi2sumintegrals} and~\eqref{eq:zeroBbgToFeynman} is included in this remainder, while
Eq.~\eqref{eq:sourcewiseVacuumZero}, valid for every $\epsilon$, excludes
an additional vacuum finite part.
Adding the local conversion in
Eq.~\eqref{eq:finiteBConversionFiniteRemainder} therefore gives
\begin{equation}
 \Delta_B\widehat c_{\rm strict}
 =\Delta_B\widehat c_{\rm rem}^{(3)}
 +\frac{2\CA}{(4\pi)^2}\,
 2(L+\gamma_E)\,\Delta_BM_0.
 \label{eq:finiteBordinaryAddback}
\end{equation}

A distinct local contribution arises from the narrow static exchanged-gluon
region. It is an overlap with the electrostatic effective theory rather than
an ultraviolet counterterm. The coefficient multiplying the local static kernel
can be written entirely in terms of the one-loop susceptibility,
\begin{align}
 \Delta_B K_E
 &=2\left[\sum_{ij}\mathcal C_{ij}(h_i-h_j)^2\right]
 T\sum_f\big[\chi_f(B)-\chi_f(0)\big]\notag\\
 &=\CA T\sum_f\big[\chi_f(B)-\chi_f(0)\big],
 \label{eq:DeltaKElocal}
\end{align}
where the second line uses $\tr \mathsf H^2=1/2$. The scalar integral
over this region is
\begin{align}
 J_\kappa(\lambda_{\rm IR})
 &\equiv\int_{|\bm k|<\kappa}\frac{\dd^3k}{(2\pi)^3}
 \frac{1}{(k^2+\lambda_{\rm IR}^2)^2}\notag\\
 &\hspace{-0.5cm}=\frac{1}{4\pi^2}
 \left[\frac{1}{\lambda_{\rm IR}}\arctan\frac{\kappa}{\lambda_{\rm IR}}
 -\frac{\kappa}{\kappa^2+\lambda_{\rm IR}^2}\right].
 \label{eq:JkappaExplicit}
\end{align}
Here $k=|\bm k|$ is the magnitude of the exchanged-gluon three-momentum,
whereas $\kappa>0$ is an auxiliary momentum-separation scale delimiting
the low-momentum region of the static exchanged-gluon contribution; it is
not another momentum variable or the renormalization scale $\Lambda$.
Extending the same integral to all three-momenta gives
\begin{equation}
 J_\infty(\lambda_{\rm IR})=\frac{1}{8\pi\lambda_{\rm IR}}.
 \label{eq:JinftyExplicit}
\end{equation}
The pointwise local term is subtracted inside the $\kappa$ ball. Conversion to
the strict dimensional-regularization prescription then adds the analytic tail
\begin{equation}
 -\Delta_B K_E\,[J_\infty(\lambda_{\rm IR})-J_\kappa(\lambda_{\rm IR})].
 \label{eq:strictDRtail}
\end{equation}
In strict matching the corresponding soft-theory integral is expanded in its
soft scales before integration and is therefore scaleless in dimensional
regularization. By contrast, the finite screened quantity
$J_\infty=1/(8\pi\lambda_{\rm IR})$ belongs to the resummed soft problem and is not part
of the strict-hard coefficient.

Combining the complete local ultraviolet subtraction, the ordinary/background
local term in Eq.~\eqref{eq:finiteBordinaryAddback}, and the strict-DR
conversion of the static overlap gives the renormalized magnetic exchange
coefficient $\Db c_{\rm strict}$. The full finite-field two-loop coefficient entering
the screening pole is therefore
\begin{equation}
 c_{B,\rm strict}
 =c_{0,\rm strict}+\Db c_{\rm strict}.
 \label{eq:cBfromDeltaCstrict}
\end{equation}
Here $c_{0,\rm strict}$ is the renormalized zero-field coefficient defined
following Eq.~\eqref{eq:zeroBassembly}.
The zero-momentum exchange contribution carries no new angular dependence
through order $g^4$; the angular dependence is generated instead by the
one-loop kinetic coefficients and the antisymmetric mixing term. Section~\ref{sec:assemblyAndTargets} combines Eq.~\eqref{eq:cBfromDeltaCstrict}
with these one-loop structures in the strict pole condition.

\section{Assembly of the screening mass and analytic numerical targets}
\label{sec:assemblyAndTargets}
The preceding sections provide all terms required by the strict pole
condition. In particular, Sec.~\ref{sec:finiteBsubtraction} supplies the
renormalized two-loop magnetic difference $\Db c_{\rm strict}$, which enters
the finite-field coefficient through
Eq.~\eqref{eq:cBfromDeltaCstrict}. We now combine this result with the
one-loop electric susceptibility, directional kinetic normalization, and
temporal-spatial mixing in Eq.~\eqref{eq:screeningFromSelfEnergy}. The
explicit finite sums, integrals, and thermodynamic derivatives used in the
numerical evaluation are then collected below.

\subsection{Hard matching coefficients and general-angle pole through \texorpdfstring{$g^4$}{g4}}
The quantities calculated in the preceding sections are precisely the four
coefficients introduced in Sec.~\ref{sec:screeningDefinition}:
$\mathcal A_B$, $z_\parallel(B)$, $z_\perp(B)$, and $\mathcal M(B)$, together
with the renormalized two-loop local coefficient $c_{B,\rm strict}$. Its finite renormalization terms are fixed by the zero-field subtraction
convention, as specified in Eq.~\eqref{eq:cBfromDeltaCstrict}. Substituting
these quantities into Eq.~\eqref{eq:hardEFTmatching} gives the complete hard
quadratic action needed through $g^4$.

A two-loop
correction to either kinetic coefficient would multiply $k^2\sim g^2T^2$ and
therefore first affect the pole at $\mathcal{O}(g^6T^2)$. Likewise, hard $\mathcal{O}(g^2)$
corrections to the spatial denominator in Eq.~\eqref{eq:selfEnergyMatrixDefinition}
affect the Schur complement only at $\mathcal{O}(g^6T^2)$.

The strict-hard screening pole is therefore the general-angle expression
already given in Eq.~\eqref{eq:screeningFromSelfEnergy}. The angle dependence
belongs to the pole of the anisotropic quadratic kernel; the local coefficient
$m_{E,{\rm loc}}^2(B)$ itself is independent of $\theta$.

For the finite-field shift, we introduce
\begin{align}
 b&\equiv\frac{eB}{T^2},
 \label{eq:bDefinition}\\
 \Delta_B\widehat{\mathcal A}
 &\equiv\frac{\Db\mathcal A}{T^2}.
 \label{eq:DeltaAhatDefinition}
\end{align}
We also have
\begin{align}
 \Delta_B z_\theta&\equiv z_\theta(B)-z_0,
 \notag\\
 \Db(\mathcal A z_\theta)
 &=\Db\mathcal A\,z_0+\mathcal A_B\Delta_B z_\theta.
 \label{eq:Aztheta}
\end{align}
Since $\mathcal M(0)=0$, the dimensionless magnetic $g^4$ pole coefficient is
\begin{align}
 \Delta_B C_\theta
 ={}&\Delta_B\widehat c_{\rm strict}
 -\frac{\Db\mathcal A\,z_0}{T^2}
 -\frac{\mathcal A_B}{T^2}\Delta_B z_\theta
 \notag\\
 &+\frac{\mathcal M^2(B)}{T^2}\sin^2\theta .
 \label{eq:dimensionlesscoeffs}
\end{align}
The magnetic hard-pole shift is
\begin{equation}
 \frac{\Db m_E^2(\theta)}{T^2}
 =g^2\Delta_B\widehat{\mathcal A}
 +g^4\Delta_B C_\theta+\mathcal{O}(g^6).
 \label{eq:hardshift}
\end{equation}
At $\theta=0,\pi/2$, Eq.~\eqref{eq:dimensionlesscoeffs} defines the parallel
and perpendicular coefficients. The coefficient at an arbitrary angle is
\begin{equation}
 \Delta_B C_\theta
 =\Delta_B C_\parallel\cos^2\theta
 +\Delta_B C_\perp\sin^2\theta.
 \label{eq:angularInterpolationExact}
\end{equation}
This angular law follows directly from the quadratic EFT operators and
requires no new multidimensional integration at intermediate angles. The
splitting is
\begin{equation}
 \Delta_B C_\perp-\Delta_B C_\parallel
 =-\frac{\mathcal A_B}{T^2}(\Delta_B z_\perp-\Delta_B z_\parallel)
 +\frac{\mathcal M^2(B)}{T^2},
 \label{eq:angularSplittingDeterministic}
\end{equation}
so the common two-loop zero-momentum coefficient drops out.

\subsection{Explicit quantities evaluated numerically}
The deterministic one-loop inputs are evaluated from the susceptibility
Eq.~\eqref{eq:chiBpub}, the finite directional representation
Eq.~\eqref{eq:Zfiniteapp}, and the mixing coefficient
Eq.~\eqref{eq:mixingCoeffPub}. For the two-loop input we factor out the
overall $g^4$. The flavor-dependent source curvature $K_f$ introduced in
Eq.~\eqref{eq:colorprojectionpub} is evaluated pointwise at finite $B$ as
\begin{align}
 K_f(h_i,h_j)
 &=\mathcal K_{qq}(h_i,h_j)+\mathcal K_{qg}(h_i,h_j)\notag\\
 &\quad+\mathcal K_{gg}(h_i,h_j).
 \label{eq:numericalKernelSum}
\end{align}
We denote its color contraction by
$\mathcal P_cK_f\equiv\sum_{ij}\mathcal C_{ij}K_f(h_i,h_j)$,
with the projection already given in Eq.~\eqref{eq:colorprojectionpub}.
The explicit kernels are given in
Eqs.~\eqref{eq:KqqExplicitMain}-\eqref{eq:KggExplicitMain}. The regulated
three-flavor exchange integral is
\begin{align}
 \mathcal I_B(\lambda_{\rm IR})
 ={}&\sum_f\frac{b_f}{2\pi}T^2
 \sum_{n,m\in\mathbb Z}
 \sum_{\ell,r\ge0}
 \notag\\
 &\times
 \int\frac{\dd p_z\dd q_z}{(2\pi)^2}
 \int\frac{\dd^2k_\perp}{(2\pi)^2}
 \mathcal P_cK_f(B;\lambda_{\rm IR}).
 \label{eq:numericalTwoLoopIntegral}
\end{align}
The multidimensional integration is performed only for the finite
three-dimensional remainder,
\begin{equation}
 \Delta_B\widehat c_{\rm rem}^{(3)}
 =-\frac{1}{2T^2}\lim_{\lambda_{\rm IR}\to0}^{\rm strict\,DR}
 \mathcal R_{\rm loc}
 \big[\mathcal I_B(\lambda_{\rm IR})-\mathcal I_0(\lambda_{\rm IR})\big],
 \label{eq:numericalCremExplicit}
\end{equation}
Equation~\eqref{eq:numericalCremExplicit} is the explicit Landau/Ritus
implementation of the ultraviolet-subtracted curvature defined in
Eq.~\eqref{eq:finiteBremFromFunctional}.
The renormalized coefficient entering Eq.~\eqref{eq:dimensionlesscoeffs} is
then
\begin{equation}
 \Delta_B\widehat c_{\rm strict}
 =\Delta_B\widehat c_{\rm rem}^{(3)}
 +\frac{2\CA}{(4\pi)^2}\,
 2(L+\gamma_E)\Delta_BM_0.
 \label{eq:numericalCstrictExplicit}
\end{equation}
The superscript on the limit in Eq.~\eqref{eq:numericalCremExplicit} denotes
the strict dimensional-regularization treatment of the static exchanged-gluon
overlap, including the analytic tail in Eq.~\eqref{eq:strictDRtail}. The
second term in Eq.~\eqref{eq:numericalCstrictExplicit} is deterministic and
is the finite local background-to-ordinary Feynman-gauge conversion derived
in Eqs.~\eqref{eq:finiteBFeynmanConversionExpanded}-~\eqref{eq:finiteBConversionFiniteRemainder}. At the reference scale
$\Lambda_0=2\pi T$, $L=-\ln2$.
In $\mathcal I_0$ the Landau measure is understood in its continuum $B\to0$
form rather than by setting $b_f=0$ inside the discrete Landau representation.

\subsection{Zero-field reduction}
As $B\to0$, $Z_\parallel=Z_\perp\to Z_0$, $\mathcal M\to0$, and
$\Delta_B c_{\rm strict}\to0$. The anisotropic quadratic action then admits
a single canonical normalization, the directional pole becomes angle
independent, $m_E^2(\theta;0)=m_E^2(0)$, and the result reduces directly to
Eq.~\eqref{eq:zeroFieldFiniteMu}.

All numerical coefficients below are quoted at the reference renormalization scale $\Lambda_0=2\pi T$. The scale dependence
of the complete magnetic coefficient at any fixed angle is
\begin{align}
 \Delta_B C_\theta(\Lambda)
 ={}&\Delta_B C_\theta(\Lambda_0)
+\frac{2\beta_0}{(4\pi)^2}
 \Delta_B\widehat{\mathcal A}
 \ln\frac{\Lambda}{\Lambda_0}
 +\mathcal{O}(g^2).
 \label{eq:RGmagnetic}
\end{align}
Here $\beta_0$ is defined in Eq.~\eqref{eq:zeroBgct}.

\subsection{Gauge-parameter independence of the hard pole}
\label{sec:gaugeIndependence}
The multidimensional source-curvature remainder is evaluated in
background-field Feynman gauge for the exchanged gluon and is converted
analytically to the ordinary linear covariant Feynman-gauge self-energy,
$\xi=1$, through Eq.~\eqref{eq:zeroBbgToFeynman}. The screening
pole discussed in this subsection is the ordinary-gauge quantity after that
conversion. Gauge-parameter independence of the pole follows from the Nielsen
identity rather than from numerical calculations at different values of
$\xi$. The relations below should therefore be read as analytic identities
for the assembled hard pole, supplemented by the ultraviolet-pole consistency
condition, not as an independent $\xi\ne1$ numerical check.

The general-covariant-gauge structure of the hard NLO gluon self-energy at
finite temperature and density has also been analyzed explicitly in
Ref.~\cite{Gorda:2023mkk}; here the gauge dependence of the assembled pole is
organized through the Nielsen identity.

The magnetic parts of $\mathcal A_B$, $z_\parallel$, $z_\perp$, and
$\mathcal M$ are one-loop quark quantities and contain no internal gluon
propagator. Hence
\begin{equation}
 \partial_\xi\Delta_B\mathcal A
 =\partial_\xi\Delta_B z_\parallel
 =\partial_\xi\Delta_B z_\perp
 =\partial_\xi\mathcal M=0,
 \label{eq:oneLoopGaugeFacts}
\end{equation}
and $z_\theta(\xi,B)=z_0(\xi)+\Delta_B z_\theta(B)$.
For the electric Schur complement the Nielsen identity may be written as~\cite{KobesKunstatterRebhan:1991,BreckenridgeLavelleSteele:1995}
\begin{equation}
 \partial_\xi\Gamma_{\rm E}(k;B)
 =2\eta_\xi(k;B)\Gamma_{\rm E}(k;B).
 \label{eq:nielsenSchur}
\end{equation}
At the order required here $\eta_\xi=g^2\eta_1(\xi)+\mathcal{O}(g^4)$. The
$\mathcal{O}(g^2)$ Nielsen kernel is generated by the gauge-fixing/ghost sector and
contains no quark loop; because the external electromagnetic field couples
only to quarks, $\eta_1$ is therefore independent of $B$. Expanding
Eq.~\eqref{eq:nielsenSchur} through $\mathcal{O}(g^4)$ gives
\begin{subequations}\label{eq:nielsenCoefficientRelations}
\begin{align}
 \partial_\xi z_0(\xi)&=2\eta_1(\xi),
 \label{eq:nielsenZ}\\
 \partial_\xi c_{B,{\rm strict}}(\xi)
 &=\mathcal A_B\,\partial_\xi z_0(\xi).
 \label{eq:nielsenC}
\end{align}
\end{subequations}
Consequently
\begin{equation}
 \partial_\xi\!\left[
 c_{B,{\rm strict}}-\mathcal A_Bz_\theta
 +\mathcal M^2\sin^2\theta\right]=0,
 \qquad
 \partial_\xi\Delta_B C_\theta=0.
 \label{eq:gaugeDifferenceIdentity}
\end{equation}
Thus the local two-loop coefficient is not separately gauge independent; its
variation cancels that of the common zero-field kinetic normalization in the
hard-pole combination.

The ultraviolet pole has the same organization. In an ordinary covariant
gauge,
\begin{align}
 W_{-1}(\xi)
 &=\frac{4\TF}{3}
 -\left(\frac{13}{6}-\frac{\xi}{2}\right)\CA,
 \notag\\[-0.2ex]
 \beta_0+W_{-1}(\xi)&=\frac{\CA}{2}(\xi+3).
 \label{eq:WminusGeneralXi}
\end{align}
Renormalization of the strict pole therefore requires
\begin{align}
 B_{-1}(B,\xi)
 &=\frac{\CA}{2}(\xi+3)\frac{\mathcal A_B}{T^2},
 \notag\\[-0.2ex]
 \Delta_BB_{-1}(\xi)
 &=\frac{\CA}{2}(\xi+3)\frac{\Delta_B\mathcal A}{T^2}.
 \label{eq:DeltaBminusGeneralXi}
\end{align}
At $B=0$ and $\xi=1$ this reduces to the directly evaluated zero-field pole
in Eq.~\eqref{eq:zeroBbarePi2}; at finite $B$ and $\xi=1$ it is precisely the
ordinary-gauge local pole displayed in
Eq.~\eqref{eq:finiteBFeynmanConversionExpanded} and canceled in the strict
pole by Eq.~\eqref{eq:finiteBStrictCounterterm}. Equation~\eqref{eq:DeltaBminusGeneralXi} is therefore a renormalization consistency
condition, not a second gauge calculation. No finite-$B$ multidimensional
integral has been evaluated at $\xi\ne1$.

\section{Numerical implementation and validation}
\label{sec:numerics}
We use $N_c=N_f=3$, massless $u,d,s$ quarks with
$(q_u,q_d,q_s)/e=(2/3,-1/3,-1/3)$, and a common chemical potential.
For $\mu/T=0$ and $1$, we study magnetic fields up to
$b=eB/T^2=4$, with $b=0$ serving as the zero-field reference.
At $b=1$ and $4$ we also evaluate the deterministic
coefficients over $0\le\mu/T\le2$ and the two-loop contribution at
$\mu/T=0,0.5,1,1.5,2$. Each two-loop integration uses
$N_{\rm scr}=12$ independently scrambled Sobol realizations with
$2^{13}$ points per cell. The proper-time quadrature uses 32 nodes,
except at $\mu/T=2$, where 64 nodes are used. For the individual
scramble estimates $I_j$, the reported QMC standard error is
\begin{align}
 \bar I
 &=\frac{1}{N_{\rm scr}}\sum_{j=1}^{N_{\rm scr}} I_j,
 \notag\\
 \sigma_{\rm QMC}
 &=\left[\frac{1}{N_{\rm scr}(N_{\rm scr}-1)}
 \sum_{j=1}^{N_{\rm scr}}(I_j-\bar I)^2\right]^{1/2}.
 \label{eq:qmcsem}
\end{align}
The one-loop susceptibility, directional kinetic coefficients, and
mixing term are evaluated by deterministic quadrature.

The production calculation uses Matsubara half-width $N=5$,
$P_{\max}/T=K_{\max}/T=64$, and $\kappa/T=1$; the auxiliary
exchanged-gluon mass is removed by the strict-DR prescription described
above. Representative tests at $b=0.125,0.25,1,$ and $4$ show a
change below $0.3\%$ in the QMC central estimate between $2^{11}$
and $2^{13}$ points per cell. Increasing the Matsubara and momentum
cutoffs changes the results at the sub-percent level, while variation
of $\kappa/T$ from $0.6$ to $1.5$ produces shifts below $0.8\%$.
The fitted residual $1/\lambda_{\rm IR}$ coefficient remains small,
and the results approach the $B\to0$ limit smoothly.

The QMC standard error measures the numerical uncertainty of the
multidimensional remainder, not a physical uncertainty. Close to zeros
of the assembled magnetic coefficient, convergence is assessed for the
remainder rather than by a relative percentage. Cutoff and
$\kappa$ variations are stability checks and are not combined in
quadrature with $\sigma_{\rm QMC}$; the ordinary/background local
conversion in Eq.~\eqref{eq:numericalCstrictExplicit} is deterministic.
The figures show central values, with QMC standard-error bars on the
discrete points in Fig.~\ref{fig:muscan}; those bars are generally
too small to distinguish at the plotted scale. The bands in
Fig.~\ref{fig:shiftpub} instead show renormalization-scale variation.

\section{Results}
\label{sec:results}
\subsection{Field-dependent strict-hard coefficients}
At $\Lambda=2\pi T$, representative field values are shown
in Table~\ref{tab:densepub}. The absolute strict-hard screening pole can be reconstructed as
\begin{align}
 \frac{m_E^2(\theta;B)}{T^2}
 ={}&g^2\frac{\mathcal A_B}{T^2}
 \notag\\
 &\hspace{-1cm}+g^4
 \left[
 C_0(\mu/T;2\pi T)+\Delta_B C_\theta(B)
 \right]
 +\mathcal{O}(g^6),
 \label{eq:absolutehardpub}
\end{align}
with $C_0(\mu/T;2\pi T)$ fixed by the zero-field result in Eq.~\eqref{eq:zeroFieldFiniteMu}.

\begin{figure}[htbp]
\centering
\includegraphics[width=.98\linewidth]{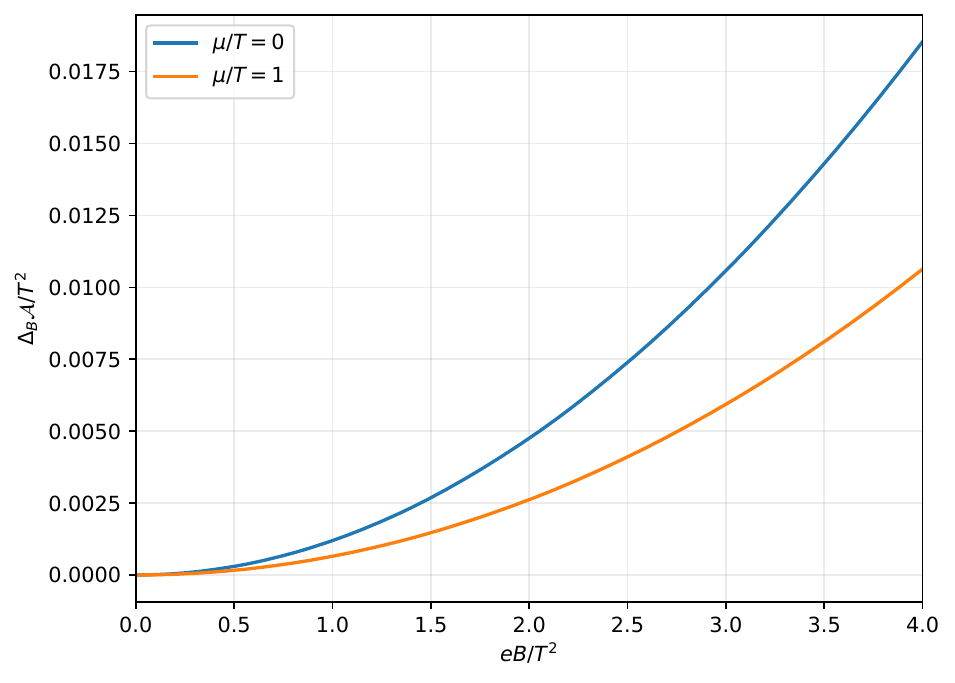}
\caption{One-loop magnetic coefficient
$\Delta_B\widehat{\mathcal A}=\Db\mathcal A/T^2$ versus $b=eB/T^2$.}
\label{fig:LOpub}
\end{figure}

\begin{figure}[tbh]
	\centering
	\includegraphics[width=.98\linewidth]{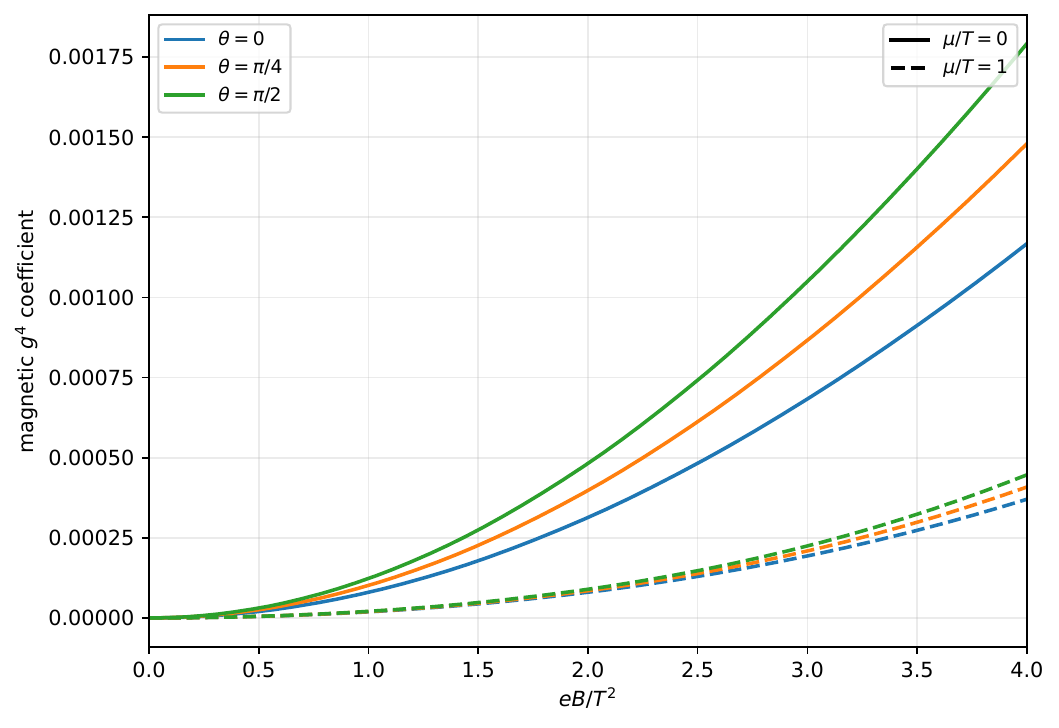}
	\caption{Magnetic $g^4$ hard coefficients versus $b=eB/T^2$ for
		$\theta=0,\pi/4,\pi/2$.}
	\label{fig:g4pub}
\end{figure}

The full field dependence of the leading magnetic coefficient is shown in Fig.~\ref{fig:LOpub}, while the directional $g^4$ coefficients are shown in
Fig.~\ref{fig:g4pub}.

\begin{widetext}
\centering
\captionof{table}{Selected renormalized strict-hard magnetic-difference
coefficients at $\Lambda=2\pi T$.}
\label{tab:densepub}
\begin{tabular}{@{}ccrrrrr@{}}\toprule
$b$ & $\mu/T$ & $10^3\Delta_B\widehat{\mathcal A}$
& $10^4\Delta_B\widehat c_{\rm strict}$
& $10^4\Delta_B C_{\theta=0}$
& $10^4\Delta_B C_{\pi/4}$
& $10^4\Delta_B C_{\pi/2}$\\\midrule
0&0&0&0&0&0&0\\
0.25&0&0.074973&0.048042&0.050361&0.063958&0.077556\\
1.00&0&1.196852&0.764285&0.800220&1.016077&1.231933\\
2.00&0&4.753947&3.004498&3.134494&3.977776&4.821058\\
4.00&0&18.542045&11.325365&11.675432&14.795754&17.916076\\
0&1&0&0&0&0&0\\
0.25&1&0.040654&0.022337&0.011669&0.012083&0.012496\\
1.00&1&0.651525&0.359468&0.190095&0.197969&0.205843\\
2.00&1&2.618925&1.462299&0.800975&0.847720&0.894465\\
4.00&1&10.637243&6.126213&3.703519&4.084216&4.464914\\
\bottomrule
\end{tabular}
\end{widetext}

The magnetic one-loop coefficient remains positive for both $\mu/T=0$
and $1$, while finite density reduces its magnitude. The magnetic $g^4$
coefficients are also positive over the displayed field range, with the
perpendicular contribution exceeding the parallel one. Both directional
coefficients decrease as the chemical potential increases.

The smallest-field points provide an internal consistency check on the smooth
$B\to0$ behavior; no polynomial extrapolation in $B$ is used in the quoted
results.
\subsection{Relative size and directional splitting}
\begin{figure}[tbh]
\centering
\includegraphics[width=.98\linewidth]{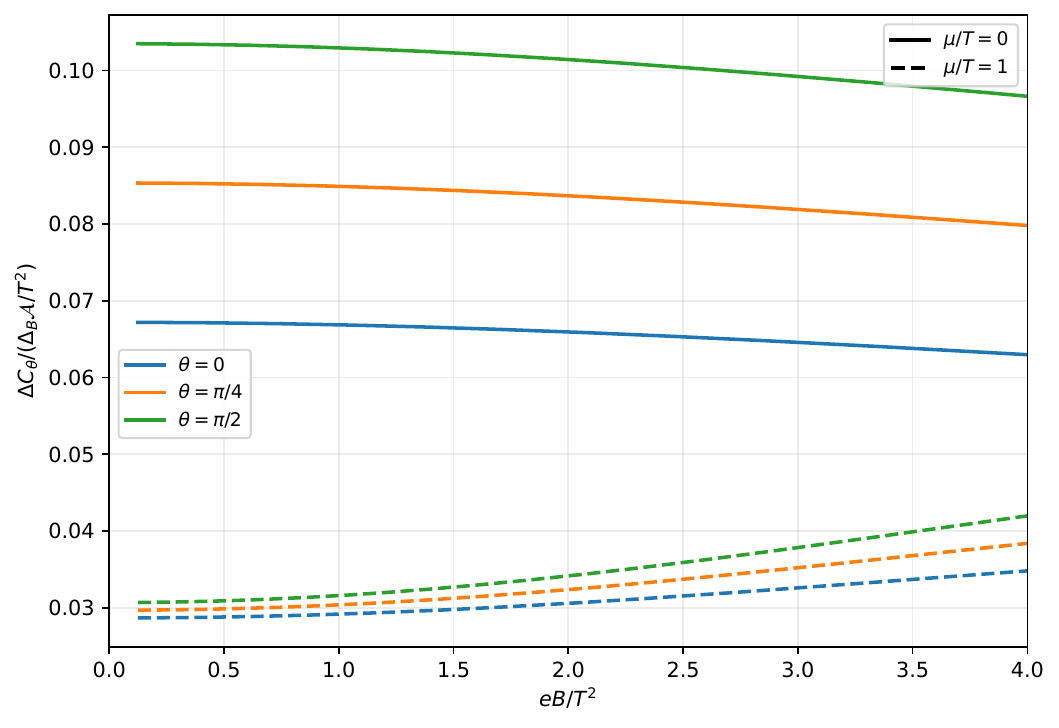}
\caption{Coupling-independent ratio
$R_\theta=\Delta_B C_\theta/\Delta_B\widehat{\mathcal A}$ versus $b=eB/T^2$.}
\label{fig:Rpub}
\end{figure}

A coupling-independent diagnostic of the perturbative correction is
\begin{equation}
 R_\theta(b,\mu/T)
 \equiv\frac{\Delta_B C_\theta}{\Delta_B\widehat{\mathcal A}}.
 \label{eq:Rpub}
\end{equation}
The ratio of the $g^4$ magnetic shift to its leading $g^2$ contribution is then
\begin{equation}
 \frac{\Db m_E^2(\theta)|_{g^4}}
 {\Db m_E^2|_{g^2}}
 =g^2R_\theta.
 \label{eq:RpubOrderRatio}
\end{equation}
A phenomenological application should insert a specified running coupling
only after the coefficient-level result has been fixed. The field dependence
of the coupling-independent ratio is displayed in Fig.~\ref{fig:Rpub}.
\subsection{Chemical-potential dependence}
The density dependence is shown in Fig.~\ref{fig:muscan}. Panels (a) and (b)
show deterministic scans, while panel (c) displays the calculated two-loop
QMC points connected by interpolation. The leading magnetic coefficient
decreases with density and eventually changes sign. The directional splitting
reverses sign at a lower chemical potential, and both magnetic $g^4$
directional coefficients also become negative at sufficiently high density.

For physical massless $u,d,s$ charges, the leading weak-field contribution to
$\mathcal M$ cancels, although higher-order field dependence remains. The
full-field numerical check at $b=1$ and $4$ gives
$\mathcal M^2/T^2\lesssim8.3\times10^{-8}$ over $0\le\mu/T\le2$.
This small mixing contribution is retained in all assembled coefficients.
We do not plot $\Delta_B C_\theta/\Delta_B\widehat{\mathcal A}$ against
$\mu/T$, because the denominator crosses zero.

The angular dependence at any fixed $(b,\mu/T)$ remains fixed analytically by
Eq.~\eqref{eq:angularInterpolationExact}; no additional angular QMC
integration is required. Since the two-loop zero-momentum contribution is a
common directional level, the difference between angles is determined by the
one-loop kinetic coefficients and the mixing term in
Eq.~\eqref{eq:angularSplittingDeterministic}. The two-loop calculation is
nevertheless essential for the absolute $g^4$ level and therefore for the
full strict-hard magnetic shift.

\subsection{\texorpdfstring{$B$}{B}-independent running-coupling illustration of the magnetic shift}
The coefficient-level results above do not require a numerical value of the
QCD coupling. For the phenomenological illustration we use a one-loop
$\overline{\rm MS}$ coupling whose scale depends on $T$ and $\mu$, but not on
the external magnetic field,
\begin{align}
 \alpha_s(\Lambda_q^2)
 &=\frac{1}{b_1\ln(\Lambda_q^2/\Lambda_{\overline{\rm MS}}^2)},
 &b_1&=\frac{11N_c-2N_f}{12\pi},
 \notag\\
 \Lambda_q&=c\sqrt{T^2+\mu^2/\pi^2}.
 \label{eq:runningBIndependent}
\end{align}
We take $\Lambda_{\overline{\rm MS}}=176\,\mathrm{MeV}$ for $N_f=3$,
$c=2\pi$ as the central scale, and $c\in[\pi,4\pi]$ for scale variation~\cite{Haque:2024gva,Haque:2014rua}.
At $T=0.4\,\mathrm{GeV}$ the central values are
$\alpha_s=0.2626$ for $\mu/T=0$ and $0.2579$ for $\mu/T=1$, independent of
$B$. Thus no magnetic dependence is inserted through the coupling itself.

Starting from the coefficient at the reference scale $\Lambda_0=2\pi T$, the
magnetic shift at a general running scale is
\begin{align}
 \frac{\Delta_Bm_E^2(\theta)}{T^2}
 ={}&g^2(\Lambda_q)\Delta_B\widehat{\mathcal A}
+g^4(\Lambda_q)\Bigg[
 \Delta_B C_\theta(2\pi T,B)
 \notag\\*
 &\hspace{-10mm}+\frac{2\beta_0}{(4\pi)^2}
 \Delta_B\widehat{\mathcal A}
 \ln\frac{\Lambda_q}{2\pi T}\Bigg]
 +\mathcal{O}(g^6).
 \label{eq:magneticShiftRunning}
\end{align}
\begin{widetext}
\centering
\includegraphics[width=.89\linewidth]{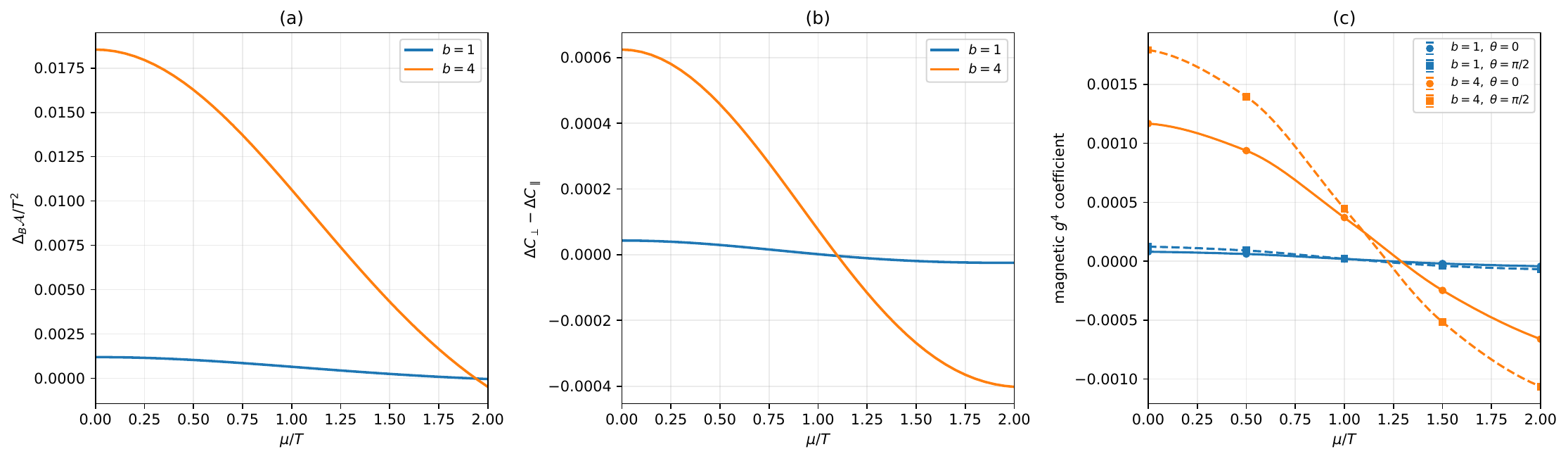}
\captionof{figure}{Chemical-potential dependence at $b=1$ and $4$: (a)
$\Delta_B\widehat{\mathcal A}$, (b)
$\Delta_B C_\perp-\Delta_B C_\parallel$, and (c)
$\Delta_B C_\parallel$ and $\Delta_B C_\perp$. Markers in panel (c) are QMC
results and the curves interpolate the calculated points.\label{fig:muscan}}
\end{widetext}

Because the same $B$-independent coupling is used for the $B$ and
$B=0$ quantities, all magnetic-field dependence through
$g^4$ resides in $\Delta_B\widehat{\mathcal A}$ and $\Delta_B C_\theta$.

Figure~\ref{fig:shiftpub} shows the resulting magnetic shift at the
representative angle $\theta=\pi/4$. The $g^4$ correction increases the
central-scale shift relative to LO, with a more pronounced effect at
$\mu/T=0$ than at $\mu/T=1$. The NLO scale-variation bands are narrower
than their LO counterparts over the displayed range, indicating reduced
residual scale sensitivity at this order. The shaded regions represent
variation of the renormalization scale, rather than QMC uncertainty.
\begin{widetext}
\centering
\includegraphics[width=.95\linewidth]{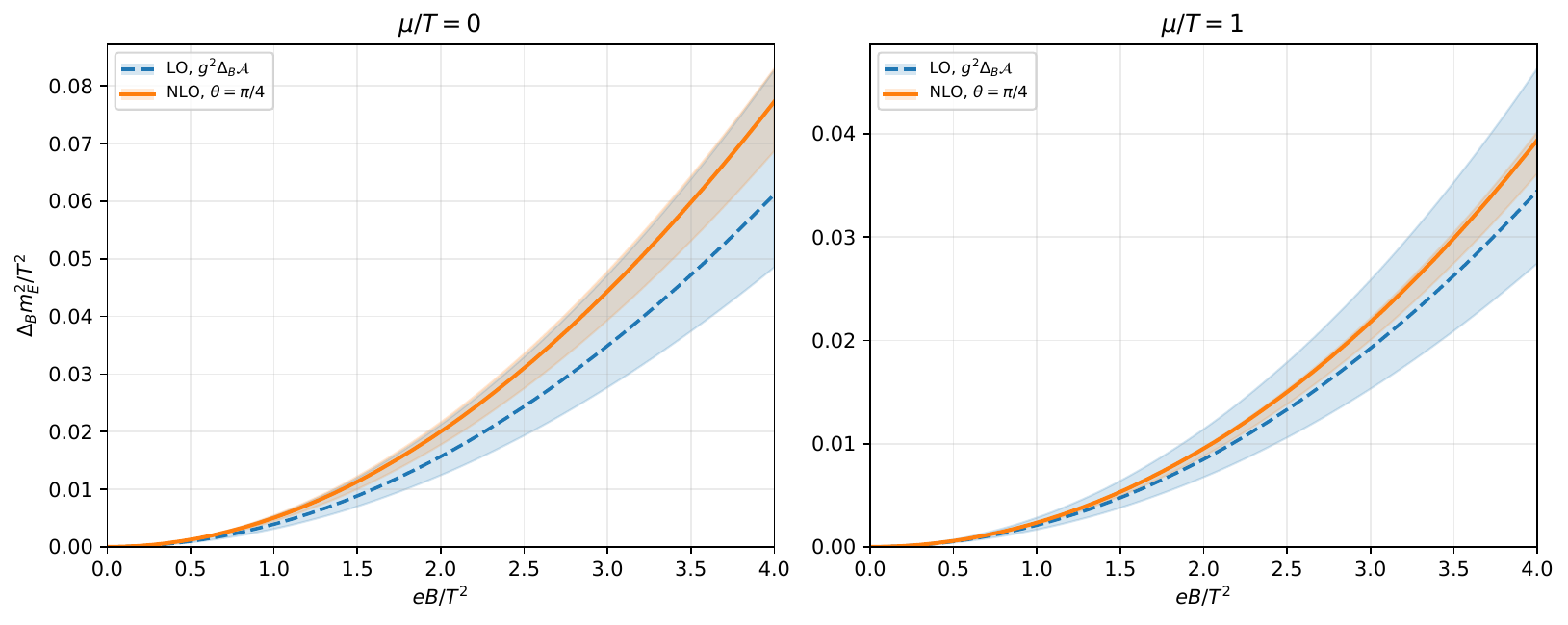}
\captionof{figure}{LO and NLO magnetic strict-hard screening-mass shift at
$\theta=\pi/4$. The bands show the scale variation $c\in[\pi,4\pi]$.}
\label{fig:shiftpub}
\end{widetext}

\FloatBarrier
\section{Summary and outlook}
\label{sec:summaryOutlook}
We have determined the anisotropic hard electric matching coefficients and
the corresponding strict-hard screening pole in hot and dense three-flavor
QCD through order $g^4$ at finite magnetic field, retaining the exact
Landau/Ritus dependence on $eB/T^2$. The calculation is a strict-hard
expansion and does not include the separately resummed soft $\mathcal{O}(g^3T^2)$
contribution to the physical long-distance Debye pole. The zero-field limit
agrees with the Braaten-Nieto result at vanishing density and with Vuorinen's
finite-density coefficient.

The finite-field two-loop exchange contribution is organized as a
background-field Feynman-gauge temporal-source curvature. We have separated
this calculational device from the ordinary covariant-Feynman self-energy
entering the screening pole and included their local conversion analytically.
The background-field magnetic difference is ultraviolet finite after the
zero-field forest subtraction because the renormalized vacuum addbacks vanish
as a $D=4-2\epsilon$ proper-time identity before the $\epsilon\to0$ limit.
The ordinary-gauge pole is generated by the local conversion and cancels,
together with its $\mathcal{O}(\epsilon)$ one-loop partner, against the corresponding
strict-pole counterterm combination. The surviving finite local term is the
$2(L+\gamma_E)\Delta_BM_0$ contribution displayed in
Eq.~\eqref{eq:finiteBordinaryAddback}. The static exchanged-gluon overlap is
treated separately with the strict dimensional-regularization prescription.

The calculation shows that for $\mu/T=0$ and $1$ the magnetic
one-loop coefficient and the magnetic $g^4$ directional coefficients are
positive throughout $0\le eB/T^2\le4$, with a substantial suppression as the
density increases. The dedicated density scan reveals a richer behavior:
$\Delta_B\widehat{\mathcal A}$ crosses zero near $\mu/T=1.91$-$1.94$ for
$b=1$-$4$, while the parallel-perpendicular splitting reverses sign already
near $\mu/T=1.03$-$1.10$. The $A_0F_{12}$ mixing term is retained in the calculation. Its leading
weak-field contribution cancels because $q_u+q_d+q_s=0$, while its full-field
contribution remains small on the plotted scales over the scanned range. Intermediate angles require no new integration because
the angular dependence follows exactly from the two endpoint coefficients.

For the illustrative mass shift we use a $B$-independent running coupling at
$\Lambda_q=c\sqrt{T^2+\mu^2/\pi^2}$. This choice keeps the fixed-order
magnetic dependence entirely in the calculated coefficients and cleanly
separates renormalization-scale variation from the randomized-QMC numerical
error.

A natural next step is to combine this hard coefficient with the magnetized
soft electrostatic sector and determine the corresponding long-distance
screening pole, including its full temporal-spatial mixing. It would also be
useful to extend the hard calculation beyond strict $\mathcal{O}(g^4)$, for example
through a systematic HTL-resummed organization that avoids double counting
between hard and soft momentum regions, and to examine finite quark-mass
effects on the arbitrary-field result.

\appendix
\section{Zero-field exchange algebra}
\label{app:zeroBalgebra}
For the normalized SU(3) source
$\mathsf H=\mathrm{diag}(1/2,-1/2,0)$, the color sums used in
Eq.~\eqref{eq:zeroBsourceDerivative} are
\begin{align*}
 \sum_f\sum_{ij}{\cal C}_{ij}h_i^2
 &=\CF\TF,\\
 \sum_f\sum_{ij}{\cal C}_{ij}h_j^2
 &=\CF\TF,
 \\
 \sum_f\sum_{ij}{\cal C}_{ij}h_ih_j
 &=-\frac{\TF}{2}(\CA-2\CF),
 \\
 \sum_f\sum_{ij}{\cal C}_{ij}\Delta h_{ij}h_i
 &=\frac{\CA\TF}{2},
 \\
 \sum_f\sum_{ij}{\cal C}_{ij}\Delta h_{ij}h_j
 &=-\frac{\CA\TF}{2},
 \\
 \sum_f\sum_{ij}{\cal C}_{ij}(\Delta h_{ij})^2
 &=\CA\TF .
\end{align*}
The temporal derivatives entering Eq.~\eqref{eq:zeroBI0reduced} are
\begin{align}
 \partial_{P_0}\mathscr H_0
 &=4(D-2)\left[
 \frac{Q_0}{P^2Q^2}
 -\frac{2P_0(P\!\cdot Q)}{(P^2)^2Q^2}
 \right],
 \label{eq:H0dP}\\
 \partial_{P_0}^2\mathscr H_0
 &=4(D-2)\Bigg[
 \frac{8P_0^2(P\!\cdot Q)}{(P^2)^3Q^2}
 \notag\\
 &\hspace{15mm}
 -\frac{2(P\!\cdot Q)}{(P^2)^2Q^2}
 -\frac{4P_0Q_0}{(P^2)^2Q^2}
 \Bigg],
 \label{eq:H0dPP}\\
 \partial_{P_0}\partial_{Q_0}\mathscr H_0
 &=4(D-2)\Bigg[
 \frac{1}{P^2Q^2}
 -\frac{2P_0^2}{(P^2)^2Q^2}
 \notag\\
 &\hspace{10mm}
 -\frac{2Q_0^2}{P^2(Q^2)^2}
 +\frac{4P_0Q_0(P\!\cdot Q)}
 {(P^2)^2(Q^2)^2}
 \Bigg].
 \label{eq:H0dPQ}
\end{align}
with the corresponding $Q_0$ derivatives obtained by $P\leftrightarrow Q$.
For the common chemical potential and $z_\mu$ defined in
Eq.~\eqref{eq:zeroBmuDefs}, the required sum-integrals are
{\allowdisplaybreaks[4]
\begin{align}
 I_1&=\frac{T^2}{12}
 \left\{1+2\epsilon\left[1+L+\frac{\zeta'(-1)}{\zeta(-1)}\right]\right\}
 +\mathcal{O}(\epsilon^2),\notag\\
 \widetilde I_1(\mu)&=-\frac{T^2}{24}
 \Big\{(1+12\bar\mu^2)+2\epsilon\big[(1+12\bar\mu^2)(1+L)\notag\\*
 &\quad+12\bigl(\zeta'(-1,z_\mu)+\zeta'(-1,z_\mu^*)\bigr)\big]\Big\}
 +\mathcal{O}(\epsilon^2),\notag\\
 I_2&=\frac1{(4\pi)^2}\left[\frac1\epsilon+2(L+\gamma_E)+\mathcal{O}(\epsilon)\right],\notag\\
 \widetilde I_2(\mu)&=\frac1{(4\pi)^2}
 \left[\frac1\epsilon+2L-\psi(z_\mu)-\psi(z_\mu^*)+\mathcal{O}(\epsilon)\right],\notag\\
 \widetilde N_2(\mu)
 &=\frac{1}{2i}\frac{\partial\widetilde I_1}{\partial\mu}
 =\frac{i\mu}{8\pi^2}+\mathcal{O}(\epsilon).
 \label{eq:zeroBsumintEval}
\end{align}}
These ingredients give the reduction in
Eq.~\eqref{eq:zeroBquarkPi2sumintegrals}.

\section{Proper-time representation of the one-loop susceptibility}
\label{app:propagator}
The one-loop electric susceptibility used in
Sec.~\ref{sec:oneLoopB} was evaluated in the exact Landau-level
representation, Eq.~\eqref{eq:chiBpub}. For completeness, in this
appendix we show how the same quantity is represented in Schwinger proper
time. This provides an independent form of the one-loop result and also
introduces the proper-time propagator used for the directional derivatives
in Appendix~\ref{app:directionalZ}.

For a uniform magnetic field, let
$b_f=|q_fB|$, $P=(P_0,\bm p)$,
$P_\parallel=(P_0,0,0,p_z)$ and
$P_\perp=(0,p_x,p_y,0)$, with
$p_\perp\equiv|\bm p_\perp|$.
With $\mathsf S=i\gamma_1\gamma_2$, the phase-stripped Euclidean quark
propagator is
\begin{equation}
 \widetilde S_{B,f}(P)
 =-i\left[
 A\slashed P_\parallel
 +C\slashed P_\perp
 -H\slashed P_\parallel\mathsf S
 \right].
 \label{eq:properTimePropagator}
\end{equation}
The proper-time weight entering these amplitudes is
\begin{equation}
 \mathcal E_f(\tau;P)
 \equiv\exp\!\left[
 -\tau(P_0^2+p_z^2)
 -\frac{p_\perp^2\tanh(b_f\tau)}{b_f}
 \right].
 \label{eq:properTimeWeight}
\end{equation}
The corresponding scalar functions, using the weight in
Eq.~\eqref{eq:properTimeWeight}, are
\begin{align}
 A&=\int_0^\infty\dd\tau\,\mathcal E_f(\tau;P),\notag\\
 C&=\int_0^\infty\dd\tau\,
 \mathrm{sech}^2(b_f\tau)\,\mathcal E_f(\tau;P),\notag\\
 H&=\operatorname{sgn}(q_fB)
 \int_0^\infty\dd\tau\,\tanh(b_f\tau)\,\mathcal E_f(\tau;P).
 \label{eq:propPTapp}
\end{align}
The propagator in Eq.~\eqref{eq:properTimePropagator} is equivalent to its
Landau-level decomposition in the Ritus representation of
Sec.~\ref{sec:oneLoopB}.
At zero transverse external momentum the
temporal vertex is diagonal in the Landau level, and the spin degeneracy is
$\alpha_\ell=2-\delta_{\ell0}$, leading to the susceptibility
Eq.~\eqref{eq:chiBpub}.

The connection between the Landau-level sum and the compact proper-time
form can be seen explicitly. When the thermal occupation factor admits a
convergent fugacity expansion,
\begin{align}
 n_F(E-\mu)\,[1-n_F(E-\mu)]
 &=\sum_{j=1}^{\infty}(-1)^{j+1}j\,\notag\\
 &\quad\times e^{-j(E-\mu)/T}.
\end{align}
The fugacity expansion is valid provided
\begin{equation}
 \operatorname{Re}(E-\mu)>0.
\end{equation}
With the analogous antiparticle expansion understood, one may use
\begin{equation}
 e^{-jE/T}
 =
 \frac{j}{2\sqrt{\pi}T}
 \int_0^\infty\frac{\dd\tau}{\tau^{3/2}}\,
 \exp\!\left[-\tau E^2-\frac{j^2}{4\tau T^2}\right].
\end{equation}
For the massless theory considered here,
$E_{f\ell}^2=p_z^2+2\ell b_f$, and the Landau-level sum is exact:
\begin{equation}
 \sum_{\ell=0}^{\infty}
 (2-\delta_{\ell0})\,e^{-2\ell b_f\tau}
 = \coth(b_f\tau).
\end{equation}
Thus the $\coth$ factor is simply the closed form of the exact Landau-level
sum in the Schwinger representation; no field expansion is involved. After
the longitudinal momentum integral, the corresponding massless QCD
coefficient may be written, wherever the fugacity series is convergent, as
\begin{align}
 \mathcal A_B
 ={}&\frac{\CA T^2}{3}
 +\frac{\TR}{\pi^2}\sum_f b_f
 \int_0^\infty\dd x\,e^{-x}
 \sum_{j=1}^\infty(-1)^{j+1}
 \notag\\
 &\times
 \cosh\frac{j\mu_f}{T}\,
 \coth\!\left(
 \frac{b_fj^2}{4xT^2}
 \right).
 \label{eq:cothQCDapp}
\end{align}
Its zero-field limit is
\begin{equation}
 \mathcal A_B\big|_{B=0}
 =\frac{\CA T^2}{3}
 +\TR\sum_f
 \left(\frac{T^2}{3}+\frac{\mu_f^2}{\pi^2}\right)
 =\mathcal A_0,
 \label{eq:cothQCDzeroB}
\end{equation}
in agreement with Eq.~\eqref{eq:A0}.

For massless quarks at nonzero chemical potential, the fugacity series is not
uniformly convergent over the full momentum range because modes with
$E_{f\ell}<|\mu_f|$ occur. Equation~\eqref{eq:cothQCDapp} is therefore not
used for the finite-density production calculation. Instead we evaluate the
exact occupation-number/Landau-level susceptibility in
Eq.~\eqref{eq:chiBpub}, which is well defined for real $\mu_f$ at $T>0$.
The two-loop magnetic calculation likewise uses the finite-density propagators
directly and not the fugacity expansion.

\section{Finite representation for the directional one-loop slopes}
\label{app:directionalZ}
For $w_n=(2n+1)\pi T-i\mu_f$, $p=p_z$, and $\rho=|\bm p_\perp|$, we define the
proper-time functions $A,C,H$ by Eq.~\eqref{eq:propPTapp}. Their derivatives
are taken under the integral. The required Dirac traces reduce to
\begin{align}
 Q_\parallel={}&w_n^2(A_p^2+H_p^2)
 -(A+pA_p)^2\notag\\
 &\hspace{3cm}-(H+pH_p)^2-\rho^2C_p^2,\notag\\
 Q_\perp=&\frac12\Big[(w_n^2-p^2)
 (A_\rho^2+H_\rho^2)-(C+\rho C_\rho)^2-C^2\Big].\!\!\!
\end{align}
We use $Q_s^{(0)}$ to denote the corresponding $B=0$ expression. Pairing the
Matsubara frequencies and using the $p\leftrightarrow-p$ symmetry yields,
for one flavor,
\begin{align}
 \Delta_B z_{s,f}^{q}
 ={}&-\frac{2\TR T}{\pi^2}
 \sum_{n=0}^\infty\Re
 \int_0^\infty\dd p
 \notag\\
 &\hspace{-1.2cm}\times
 \int_0^\infty \rho\,\dd \rho\,
 \big[
 Q_s(w_n,p,\rho;b_f)
 -Q_s^{(0)}(w_n,p,\rho)
 \big].
 \label{eq:Zfiniteapp}
\end{align}
The total magnetic difference is $\Delta_B z_s=\sum_f\Delta_B z_{s,f}^{q}$.
With a common regulator for the finite- and zero-field contributions in
Eq.~\eqref{eq:Zfiniteapp}, the subtracted expression is ultraviolet finite
and can be evaluated directly for both $s=\parallel$ and $s=\perp$.

The mixing coefficient entering the transverse result follows from
the matter pressure per color,
\begin{align}
 p_f
 ={}&\frac{b_fT}{2\pi^2}
 \sum_{\ell=0}^\infty\alpha_\ell
 \int_0^\infty\dd p_z
 \notag\\
 &\times
 \Big[
 \ln(1+e^{-(E_{f\ell}-\mu_f)/T})
 \notag\\
 &\qquad+
 \ln(1+e^{-(E_{f\ell}+\mu_f)/T})
 \Big].
\end{align}
Here $n_f=\partial p_f/\partial\mu_f$, and $\mathcal H_f$ follows from
Eq.~\eqref{eq:mixingHfDefinition}.


\begin{thebibliography}{99}
\bibitem{Braaten:1995jr}
E.~Braaten and A.~Nieto,
Phys. Rev. D \textbf{53}, 3421-3437 (1996),
\href{https://doi.org/10.1103/PhysRevD.53.3421}{doi:10.1103/PhysRevD.53.3421}.


\bibitem{Vuorinen:2003fs}
A.~Vuorinen,
Phys. Rev. D \textbf{68}, 054017 (2003),
\href{https://doi.org/10.1103/PhysRevD.68.054017}{doi:10.1103/PhysRevD.68.054017}.

\bibitem{Kajantie:1995dw}
K.~Kajantie, M.~Laine, K.~Rummukainen and M.~E.~Shaposhnikov,
Nucl. Phys. B \textbf{458}, 90-136 (1996),
\href{https://doi.org/10.1016/0550-3213(95)00549-8}{doi:10.1016/0550-3213(95)00549-8},
\href{https://arxiv.org/abs/hep-ph/9508379}{arXiv:hep-ph/9508379 [hep-ph]}.

\bibitem{Laine:2005ai}
M.~Laine and Y.~Schr\"oder,
JHEP \textbf{03}, 067 (2005),
\href{https://doi.org/10.1088/1126-6708/2005/03/067}{doi:10.1088/1126-6708/2005/03/067},
\href{https://arxiv.org/abs/hep-ph/0503061}{arXiv:hep-ph/0503061 [hep-ph]}.

\bibitem{Rebhan:1993az}
A.~K.~Rebhan,
Phys. Rev. D \textbf{48}, R3967-R3970 (1993),
\href{https://doi.org/10.1103/PhysRevD.48.R3967}{doi:10.1103/PhysRevD.48.R3967}.

\bibitem{Kapusta:1992fm}
J. I. Kapusta,
``Screening of static QED electric fields in hot QCD,''
\emph{Phys. Rev. D} \textbf{46}, 4749 (1992),
\href{https://doi.org/10.1103/PhysRevD.46.4749}{doi:10.1103/PhysRevD.46.4749}.

\bibitem{Blaizot:1995kg}
J. P. Blaizot, E. Iancu and R. R. Parwani,
``Screening of static electromagnetic fields in hot QED plasmas,''
\emph{Phys. Rev. D} \textbf{52}, 2543 (1995),
\href{https://doi.org/10.1103/PhysRevD.52.2543}{doi:10.1103/PhysRevD.52.2543},
\href{https://arxiv.org/abs/hep-ph/9504408}{arXiv:hep-ph/9504408}.

\bibitem{Arnold:1995bh}
P. B. Arnold and L. G. Yaffe,
``The non-Abelian Debye screening length beyond leading order,''
\emph{Phys. Rev. D} \textbf{52}, 7208 (1995),
\href{https://doi.org/10.1103/PhysRevD.52.7208}{doi:10.1103/PhysRevD.52.7208},
\href{https://arxiv.org/abs/hep-ph/9508280}{arXiv:hep-ph/9508280}.

\bibitem{Ghisoiu:2015uza}
I. Ghisoiu, J. M\"oller and Y. Schr\"oder,
``Debye screening mass of hot Yang-Mills theory to three-loop order,''
\emph{JHEP} \textbf{11}, 121 (2015),
\href{https://doi.org/10.1007/JHEP11(2015)121}{doi:10.1007/JHEP11(2015)121},
\href{https://arxiv.org/abs/1509.08727}{arXiv:1509.08727}.

\bibitem{Matsui:1986dk}
T.~Matsui and H.~Satz,
Phys. Lett. B \textbf{178}, 416-422 (1986),
\href{https://doi.org/10.1016/0370-2693(86)91404-8}
{doi:10.1016/0370-2693(86)91404-8}.


\bibitem{Karsch:1987pv}
F.~Karsch, M.~T.~Mehr and H.~Satz,
Z. Phys. C \textbf{37}, 617 (1988),
\href{https://doi.org/10.1007/BF01549722}
{doi:10.1007/BF01549722}.

\bibitem{Skokov:2009qp}
V.~Skokov, A.~Y.~Illarionov and V.~Toneev,
``Estimate of the magnetic field strength in heavy-ion collisions,''
Int. J. Mod. Phys. A \textbf{24}, 5925-5932 (2009),
\href{https://doi.org/10.1142/S0217751X09047570}{doi:10.1142/S0217751X09047570},
\href{https://arxiv.org/abs/0907.1396}{arXiv:0907.1396 [nucl-th]}.

\bibitem{Deng:2012pc}
W.-T.~Deng and X.-G.~Huang,
``Event-by-event generation of electromagnetic fields in heavy-ion collisions,''
Phys. Rev. C \textbf{85}, 044907 (2012),
\href{https://doi.org/10.1103/PhysRevC.85.044907}{doi:10.1103/PhysRevC.85.044907},
\href{https://arxiv.org/abs/1201.5108}{arXiv:1201.5108 [nucl-th]}.

\bibitem{Tuchin:2013ie}
K.~Tuchin,
Adv. High Energy Phys. \textbf{2013}, 490495 (2013),
\href{https://doi.org/10.1155/2013/490495}{doi:10.1155/2013/490495},
\href{https://arxiv.org/abs/1301.0099}{arXiv:1301.0099 [hep-ph]}.

\bibitem{Karmakar:2019tdp}
B.~Karmakar, R.~Ghosh, A.~Bandyopadhyay, N.~Haque and M.~G.~Mustafa,
Phys. Rev. D \textbf{99}, no.9, 094002 (2019),
\href{https://doi.org/10.1103/PhysRevD.99.094002}{doi:10.1103/PhysRevD.99.094002},
\href{https://arxiv.org/abs/1902.02607}{arXiv:1902.02607 [hep-ph]}.

\bibitem{Karmakar:2018aig}
B.~Karmakar, A.~Bandyopadhyay, N.~Haque and M.~G.~Mustafa,
Eur. Phys. J. C \textbf{79}, no.8, 658 (2019),
\href{https://doi.org/10.1140/epjc/s10052-019-7154-0}{doi:10.1140/epjc/s10052-019-7154-0},
\href{https://arxiv.org/abs/1804.11336}{arXiv:1804.11336 [hep-ph]}.

\bibitem{Birse:2003qp}
M. C. Birse, C. W. Kao and G. C. Nayak,
``Screening and antiscreening in anisotropic QED and QCD plasmas,''
\emph{Phys. Lett. B} \textbf{570}, 171 (2003),
\href{https://doi.org/10.1016/j.physletb.2003.08.007}{doi:10.1016/j.physletb.2003.08.007},
\href{https://arxiv.org/abs/hep-ph/0304209}{arXiv:hep-ph/0304209}.

\bibitem{Bonati:2014ksa}
C.~Bonati, M.~D'Elia, M.~Mariti, M.~Mesiti, F.~Negro and F.~Sanfilippo,
Phys. Rev. D \textbf{89}, no.11, 114502 (2014),
\href{https://doi.org/10.1103/PhysRevD.89.114502}{doi:10.1103/PhysRevD.89.114502},
\href{https://arxiv.org/abs/1403.6094}{arXiv:1403.6094 [hep-lat]}.

\bibitem{Bonati:2017uvz}
C. Bonati, M. D'Elia, M. Mariti, M. Mesiti, F. Negro, A. Rucci and F. Sanfilippo,
``Screening masses in strong external magnetic fields,''
\emph{Phys. Rev. D} \textbf{95}, 074515 (2017),
\href{https://doi.org/10.1103/PhysRevD.95.074515}{doi:10.1103/PhysRevD.95.074515},
\href{https://arxiv.org/abs/1703.00842}{arXiv:1703.00842}.


\bibitem{Sebastian:2023tlw}
J.~Sebastian, L.~Thakur, H.~Mishra and N.~Haque,
Phys. Rev. D \textbf{108}, no.9, 094001 (2023),
\href{https://doi.org/10.1103/PhysRevD.108.094001}{doi:10.1103/PhysRevD.108.094001},
\href{https://arxiv.org/abs/2308.04410}{arXiv:2308.04410 [hep-ph]}.


\bibitem{Ayala:2019akk}
A.~Ayala, J.~D.~Casta{\~n}o-Yepes, M.~Loewe and E.~Mu{\~n}oz,
Phys. Rev. D \textbf{101}, no.3, 036016 (2020),
\href{https://doi.org/10.1103/PhysRevD.101.036016}{doi:10.1103/PhysRevD.101.036016}.


\bibitem{Alexandre:2000jc}
J.~Alexandre,
Phys. Rev. D \textbf{63}, 073010 (2001),
\href{https://doi.org/10.1103/PhysRevD.63.073010}{doi:10.1103/PhysRevD.63.073010},
\href{https://arxiv.org/abs/hep-th/0009204}{arXiv:hep-th/0009204 [hep-th]}.


\bibitem{Bandyopadhyay:2017cle}
A.~Bandyopadhyay, B.~Karmakar, N.~Haque and M.~G.~Mustafa,
Phys. Rev. D \textbf{100}, no.3, 034031 (2019),
\href{https://doi.org/10.1103/PhysRevD.100.034031}{doi:10.1103/PhysRevD.100.034031}.

\bibitem{Fukushima:2011nu}
K.~Fukushima,
Phys. Rev. D \textbf{83}, 111501 (2011),
\href{https://doi.org/10.1103/PhysRevD.83.111501}{doi:10.1103/PhysRevD.83.111501},
\href{https://arxiv.org/abs/1103.4430}{arXiv:1103.4430 [hep-ph]}.


\bibitem{Fraga:2023cef}
E.~S.~Fraga, L.~F.~Palhares and T.~E.~Restrepo,
Phys. Rev. D \textbf{108}, no.3, 034026 (2023),
\href{https://doi.org/10.1103/PhysRevD.108.034026}{doi:10.1103/PhysRevD.108.034026}.

\bibitem{Maezawa:2010vj}
Y. Maezawa \textit{et al.} [WHOT-QCD Collaboration],
``Electric and magnetic screening masses at finite temperature from generalized Polyakov-line correlations in two-flavor lattice QCD,''
\emph{Phys. Rev. D} \textbf{81}, 091501 (2010),
\href{https://doi.org/10.1103/PhysRevD.81.091501}{doi:10.1103/PhysRevD.81.091501},
\href{https://arxiv.org/abs/1003.1361}{arXiv:1003.1361}.

\bibitem{Borsanyi:2015yka}
S. Bors\'anyi, Z. Fodor, S. D. Katz, A. P\'asztor, K. K. Szab\'o and C. T\"or\"ok,
``Static $\overline{Q}Q$ pair free energy and screening masses from correlators of Polyakov loops: continuum extrapolated lattice results at the QCD physical point,''
\emph{JHEP} \textbf{04}, 138 (2015),
\href{https://doi.org/10.1007/JHEP04(2015)138}{doi:10.1007/JHEP04(2015)138},
\href{https://arxiv.org/abs/1501.02173}{arXiv:1501.02173}.

\bibitem{Bazavov:2020teh}
A. Bazavov and J. H. Weber,
``Color screening in quantum chromodynamics,''
\emph{Prog. Part. Nucl. Phys.} \textbf{116}, 103823 (2021),
\href{https://doi.org/10.1016/j.ppnp.2020.103823}{doi:10.1016/j.ppnp.2020.103823},
\href{https://arxiv.org/abs/2010.01873}{arXiv:2010.01873}.


\bibitem{Wang:2021eud}
X.~Wang and I.~Shovkovy,
Eur. Phys. J. C \textbf{81}, no.10, 901 (2021),
\href{https://doi.org/10.1140/epjc/s10052-021-09650-3}{doi:10.1140/epjc/s10052-021-09650-3}.

\bibitem{Gorda:2023mkk}
T.~Gorda, R.~Paatelainen, S.~S\"appi, and K.~Sepp\"anen,
JHEP \textbf{08}, 021 (2023),
\href{https://doi.org/10.1007/JHEP08(2023)021}{doi:10.1007/JHEP08(2023)021},
\href{https://arxiv.org/abs/2304.09187}{arXiv:2304.09187 [hep-ph]}.

\bibitem{KobesKunstatterRebhan:1991}
R.~Kobes, G.~Kunstatter, and A.~K.~Rebhan,
Nucl. Phys. B \textbf{355}, 1-37 (1991),
\href{https://doi.org/10.1016/0550-3213(91)90300-M}{doi:10.1016/0550-3213(91)90300-M}.

\bibitem{BreckenridgeLavelleSteele:1995}
J.~C.~Breckenridge, M.~J.~Lavelle, and T.~G.~Steele,
Z. Phys. C \textbf{65}, 155-164 (1995),
\href{https://doi.org/10.1007/BF01571316}{doi:10.1007/BF01571316},
\href{https://arxiv.org/abs/hep-th/9407028}{arXiv:hep-th/9407028}.


\bibitem{Haque:2024gva}
N.~Haque and M.~G.~Mustafa,
Prog. Part. Nucl. Phys. \textbf{140}, 104136 (2025),
\href{https://doi.org/10.1016/j.ppnp.2024.104136}{doi:10.1016/j.ppnp.2024.104136},
\href{https://arxiv.org/abs/2404.08734}{arXiv:2404.08734 [hep-ph]}.

\bibitem{Haque:2014rua}
N.~Haque, A.~Bandyopadhyay, J.~O.~Andersen, M.~G.~Mustafa, M.~Strickland and N.~Su,
JHEP \textbf{05}, 027 (2014),
\href{https://doi.org/10.1007/JHEP05(2014)027}{doi:10.1007/JHEP05(2014)027},
\href{https://arxiv.org/abs/1402.6907}{arXiv:1402.6907 [hep-ph]}.

\end{thebibliography}
\end{document}